\documentclass{article}
\usepackage[utf8]{inputenc}
\usepackage{color}
\usepackage{epsfig}
\usepackage{amsfonts}
\usepackage{floatrow}
\newfloatcommand{capbtabbox}{table}[][\FBwidth]
\usepackage{blindtext}
\usepackage{amsmath}
\usepackage{amssymb}
\usepackage{rotating}
\usepackage{framed}
\usepackage{hyperref}
\usepackage{graphicx}
\usepackage{multirow}
\usepackage{chngpage}
\usepackage{floatrow}
\usepackage{blindtext}

\title{Connecting Instrumental Variable methods for causal inference to the Estimand Framework}
\author{Jack Bowden$^{1,2}$, Bj{\"o}rn Bornkamp$^{3}$, Ekkehard Glimm$^{3,4}$\\ 
 \& Frank Bretz$^{3,5}$}

\date{}

\begin{document}

\maketitle

\begin{flushleft}
{\it$^{1}$Exeter Diabetes Group (ExCEED), College of Medicine and Health, University of Exeter, Exeter, U.K.}\\
{\it$^{2}$MRC Integrative Epidemiology Unit at the University of Bristol, U.K.}\\
{\it$^{3}$Novartis Pharma AG, CH-4002 Basel, Switzerland.}\\
{\it$^{4}$Institute for Biometry and Medical Infromatics, Medical Faculty, University of Magdeburg, Germany}\\
{\it$^{5}$Section for Medical Statistics, Medical University of Vienna, Vienna, Austria.}\\
\end{flushleft}
\hspace{4cm}
\begin{center}
$^{*}$Address for correspondence: \\
Jack Bowden, University of Exeter College of Medicine and Health\\
RILD Building, RD\&E Hospital, \\
Barrack Road, Exeter, Devon, EX2 5DW\\
j.bowden2@exeter.ac.uk
\end{center}

{\bf Abstract}\\
\\
Causal inference methods are gaining increasing prominence in pharmaceutical drug development in light of the recently published addendum on estimands and sensitivity analysis in clinical trials to the E9 guideline of the International Council for Harmonisation.  The E9 addendum emphasises the need to account for post-randomization or `intercurrent' events that can potentially influence the interpretation of a treatment effect estimate at a trial's conclusion. Instrumental Variables (IV) methods have been used extensively in economics, epidemiology and academic clinical studies for `causal inference', but less so in the pharmaceutical industry setting until now. In this tutorial paper we review the basic tools for causal inference, including graphical diagrams and potential outcomes, as well as several conceptual frameworks that an IV analysis can sit within. We discuss in detail how to map these approaches to the Treatment Policy, Principal Stratum and Hypothetical `estimand strategies' introduced in the E9 addendum, and provide details of their implementation using standard regression models. Specific attention is given to discussing the assumptions each estimation strategy relies on in order to be consistent, the extent to which they can be empirically tested and sensitivity analyses in which specific assumptions can be relaxed. We finish by applying the methods described to simulated data closely matching two recent pharmaceutical trials to further motivate and clarify the ideas.

\newpage
\section{Introduction}

What is the effect of treatment on an individual patient's health in a trial?  In order to answer this question we would need to measure how a patient's outcome would have changed if they had been given the treatment compared to if they had not. Counter-factual contrasts like this are a popular vehicle for defining a causal effect, but they illustrate the `fundamental problem' of causal inference: It is not possible to directly observe the outcome under both treatment choices for a single individual under identical conditions \cite{holland1986}. In specific settings, such as a crossover trial, or in ophthalmology - where one eye might be treated and the other not - one can come close to this ideal, but specific assumptions must still be made. In an idealised randomized controlled trial (RCT) of adequate size in which all patients adhere to their assigned treatment regimen until their final outcome is observed, the act of randomization ensures that patients on each treatment arm will be sufficiently balanced with respect to all patient characteristics. This provides a solid rationale for attributing any difference in outcomes between the two treatment groups at the end of the trial to the treatment itself, and for unbiasedly estimating this difference (the treatment effect) by comparing patient outcomes across randomized groups. In the field of epidemiology this is referred to as a `causal' effect estimate and the concept of an idealised or `target' RCT is routinely invoked in the observational sciences to explain the underlying notion of causality \cite{robins1986,robins2000,smith2003}. In spite of this, the word `causal' is rarely - if ever - used within the clinical trials arena. Paradoxically, some high profile journals think it should {\it only} be used within an RCT \cite{hernan2018}.\\ 
\\
When running a RCT in practice, a certain fraction of the patients may not adhere to the assigned treatment regimen, by either skipping a dose, reducing it, or stopping altogether. This could be due to an inability to tolerate a treatment or a lack of efficacy. Such departures are very much part of the standard treatment process. At the same time, understanding what the trial results would have looked like if such departures had not occurred may in some situations also be of interest (in particular when departures can be prevented in a practical setting). Naively accounting for non-adherence via so called As Treated, Per Protocol or Responder analyses can be misleading, whenever patient characteristics that predict non-adherence to treatment also influence the outcome, because of the re-introduction of confounding into trial data \cite{greenland2000,greenland2008}. Unfortunately, an intention-to-treat (ITT) analysis of patients according to the original randomized groups (irrespective of non-adherence), is not a catch-all solution either \cite{hernan2012}. In particular, the question remains whether estimating an effect in accordance with the ITT principle always represents the effect of greatest relevance to regulatory and clinical decision making. In some settings we may want to know the likely benefit achieved when the treatment is given in the routine care setting, which naturally encompasses a degree of non-adherence within it. Alternatively, we may instead be interested in estimating the effect of treatment for only those who do adhere \cite{shrier2014}. So called `Instrumental Variable' methods can be used to go beyond the ITT analysis to answer questions of this nature \cite{cuzick1997}, and are the major focus of this paper.\\
\\
An important document relating to this discussion is the addendum on estimands and sensitivity analysis in clinical trials to the E9 guideline of the International Council for Harmonisation \cite{ICHE9}, which we refer to as 'E9 addendum'.  It defines any such post randomization event that potentially influences the interpretation of a treatment effect estimate at the end of the trial as an `intercurrent event'. It stresses the need to pro-actively address the issue of intercurrent events in the design, analysis and reporting of studies. The working principles described in the E9 addendum are referred to as the `Estimand Framework' \cite{akacha2017,akacha2017a}, because it accentuates the importance of defining the specific target (or estimand) that is to be estimated, see also \cite{qu2020general,guizzaro2020use}. 

Despite early work on the application of Instrumental Variable (IV) methods to RCTs \cite{permutt1989,sommer1991} they have been seldom used in practice. In this paper we consider the application of IV methods for pursuing three distinct estimand strategies described in the E9 Addendum:
\begin{enumerate}
\item{The {\it Treatment Policy } strategy is one for which an intercurrent event becomes a fundamental attribute of the treatment.  A Treatment Policy estimand is therefore  the effect of treatment in patients irrespective of whether they experienced the intercurrent event or not.}
\item{The {\it  Principal Stratum}  strategies target the effect of treatment within a subgroup of patients who would not have experienced the intercurrent event under assignment to one or more  treatments.}
\item{The  {\it Hypothetical} strategies target the effect of treatment under a hypothetical scenario that the intercurrent event would not have occurred.}
\end{enumerate}
We restrict our attention to trials with either a continuous or binary outcome and focus on estimating treatment effects on the mean- or risk-difference scale. In this context we review the conceptual framework of Principal Stratification \cite{angrist1996,frangakis2002} for targeting Principal Stratum estimands, and Structural Mean Models \cite{goetghebeur1997,fischer2011} for targeting Hypothetical estimands. We use causal diagrams and potential outcomes to explain their rationale and the estimand they target.\\
\\ 
The aim of this tutorial is to bridge the gap between statisticians engaged in bio-pharmaceutical research and the causal inference community. In what follows, we provide an accessible introduction of IV methods that have been used extensively in epidemiology and academic clinical studies for causal inference, but less so in the pharmaceutical industry setting until now. Accordingly, the paper is organized as follows. In Section~\ref{s:ex}, we introduce several RCT examples in a contemporary industry setting to motivate the subsequent sections. In Sections~\ref{s:ci}, \ref{s:est} and \ref{s:ass}, we explain IV methods when the intercurrent event is all or nothing adherence to treatment, as classically assumed within the academic IV literature. To pharmaceutical statisticians, this setting may appear to have only narrow relevance to the problems encountered in clinical trials, but in order to bring the communities together, this is a necessary starting point for a common language. More specifically, we introduce basic causal inference notation, connect the IV methods to the Estimand Framework from \cite{ICHE9} and discuss the underlying assumptions each estimation strategy relies on.  In Section~\ref{s:exrev}, we discuss in detail two clinical trial examples to show how the classical setting can be adapted to better serve this need, using a combination of features from the Treatment Policy, Principal Stratum and Hypothetical estimand strategies. Concluding remarks are given in Section~\ref{s:disc}. \\
\\
This is by no means the first review article of this nature, see for example \cite{greenland2000,bellamy2007,lipkovich2020}. Our contribution attempts to provide an accessible account of the tools and techniques for IV estimation used in the classical causal inference community for a pharmaceutical statistics audience within the context of the Estimand Framework introduced in the E9 addendum \cite{ICHE9}. To this end, the paper is organized such that the main body contains a non-technical description of the methods. This is interspersed with figures containing additional modelling details and graphical explanations for the interested reader. R code to implement all of the methods described is provided in {\it Supplementary Material}.

\section{Clinical trial examples from a contemporary industry setting}
\label{s:ex}

In this tutorial we discuss how IV-based estimand strategies can be used to adjust for intercurrent events of relevance to realistic contemporary clinical trials. Specifically, rather than simply assuming non-compliance with the randomized treatment from the start, as classically done within the academic IV literature, we will assume that intercurrent events can also be some other unplanned disturbance of the trial and can sit between the initiation of treatment and measurement of the final outcome. Possible examples include:
\begin{enumerate}
    \item{The presence of disease progression to stage 4 cancer after completion of chemotherapy regimen in a trial measuring overall survival at 5 years;}
    \item{The absence of antibodies to a virus three weeks after being administered a vaccine in a trial measuring re-infection within 2 years;}
    \item{The presence of `relapse' following treatment in patients with secondary progressive multiple sclerosis, before the final disability assessment at 3 months.}
\end{enumerate}
The last example is motivated by the  EXPAND trial (NCT01665144 on ClinicalTrials.gov) and in recent work \cite{magnusson2019}, Bayesian methods were proposed for this trial to quantify  Principal Stratum estimands within patient subgroups defined by relapse status.\\ 
\\ 
To make ideas concrete, we will consider a randomized placebo controlled trial,  inspired by the Canakinumab Anti-inflammatory Thrombosis Outcome Study (CANTOS) trial \cite{ridker2017,bornkamp2019}. It sought to evaluate whether Canakinumab, a monoclonal antibody which acts to reduce inflammation, was effective in reducing the risk of a major cardiac event in approximately 10,000 patients. For our hypothetical trial, we assume that the intercurrent event of interest is measured by a relevant biomarker 1 month after initiation of treatment. The trial outcome is death at or before 3 years. The experimental treatment is hypothesised to work directly through the biomarker so that, if a treated patient does not respond, we believe that the drug has failed to work as planned. Likewise, if a patient does not receive the treatment but nevertheless has a positive biomarker response after 1 month, we may believe that their future health outcomes have been improved or worsened in line with those who took and responded to treatment. \\
\\
In this setting a naive `responder' analysis would quantify the association between biomarker response and mortality by conditioning on the patients' observed biomarker response. This does not have an automatic causal interpretation in its own right, because it directly conditions on the observed (post-randomization) intercurrent event, which could be confounded with the study outcome. We will subsequently show how to use IV methods to obtain a fair estimate for the effectiveness of the treatment in (a) principled patient sub-groups defined by biomarker response, or (b) to estimate the causal effect of biomarker response directly, and to achieve (a) or (b) without making the assumption that all confounders of biomarker response and the outcome can be measured and adjusted for.\\
\\
We also consider a second class of intercurrent events commonly encountered in industry trials, that of `general' non-adherence. A recent example described in Qu et. al \cite{qu2020general} is the IMAGINE-3 study, which compared the use of basal versus glargine insulin treatment for controlling HbA1c levels in type I diabetic patients. Of the 1112 patients initially randomized to treatment and who took at least one dose, approximately 76\% of the basal insulin and 82\% of the glargine insulin medication adhered to the full 52-week treatment course. In this example, and others like it,  it would not be reasonable to assume that adherence lies directly on the causal pathway (i.e. that adherence alone is a near-perfect predictor of treatment effect such as biomarker response was assumed to be in the previous example). In this setting, a naive `per-protocol' analysis would be to compare outcomes across randomised groups using only those patients who adhered to the treatment given. Again, this lacks an immediate causal interpretation because adherence is a post-randomization event. Qu et al \cite{qu2020general} propose a general framework for estimating treatment effects in `principled' patient subgroups defined by adherence, based on the assumption that all confounders of adherence and outcome are measured. Qu et al's framework extends to the case of missing outcome data, under the missing at random assumption. We will show that it is possible to recover the same causal effects estimated by Qu et al without assuming knowledge of all confounders, as long as the outcomes for all patients are observed.

\section{Tools for causal inference in clinical trials}
\label{s:ci}

Let $R$, $T$ and $Y$ represent the data collected on each patient within a generic two-arm randomized controlled trial. Here, $R$ denotes randomization to the experimental treatment ($R$ = 1) or control ($R$ = 0). The variable $T$ denotes whether a patient subsequently receives the treatment ($T=1$) or control ($T=0$) therapy. Finally, $Y$ denotes the continuous or binary patient outcome of interest, on which the two randomized groups are to be compared. \\
\\
We are interested in the scenario whereby post-randomization (in this case whether a patient takes the treatment they were randomized to receive) may influence the interpretation or the estimation of the treatment effect.In many pharmaceutical trial settings, especially when both the patient and clinician are blinded, it may of course be impossible for the control arm patient to receive the experimental treatment. Furthermore, typically the definition of the intercurrent events is more nuanced in a clinical trial. For example, treatment discontinuation or intake of additional rescue medication might instead be used, with each one being handled with different estimand strategies. We include this simplistic intercurrent event definition for the purposes of clarity and generality, especially for those from an academic or epidemiological background. We then consider more nuanced cases in Section 6.

\subsection{Causal diagrams}

\begin{figure}[htbp]
    \centering
\includegraphics[width=0.85\textwidth,clip]{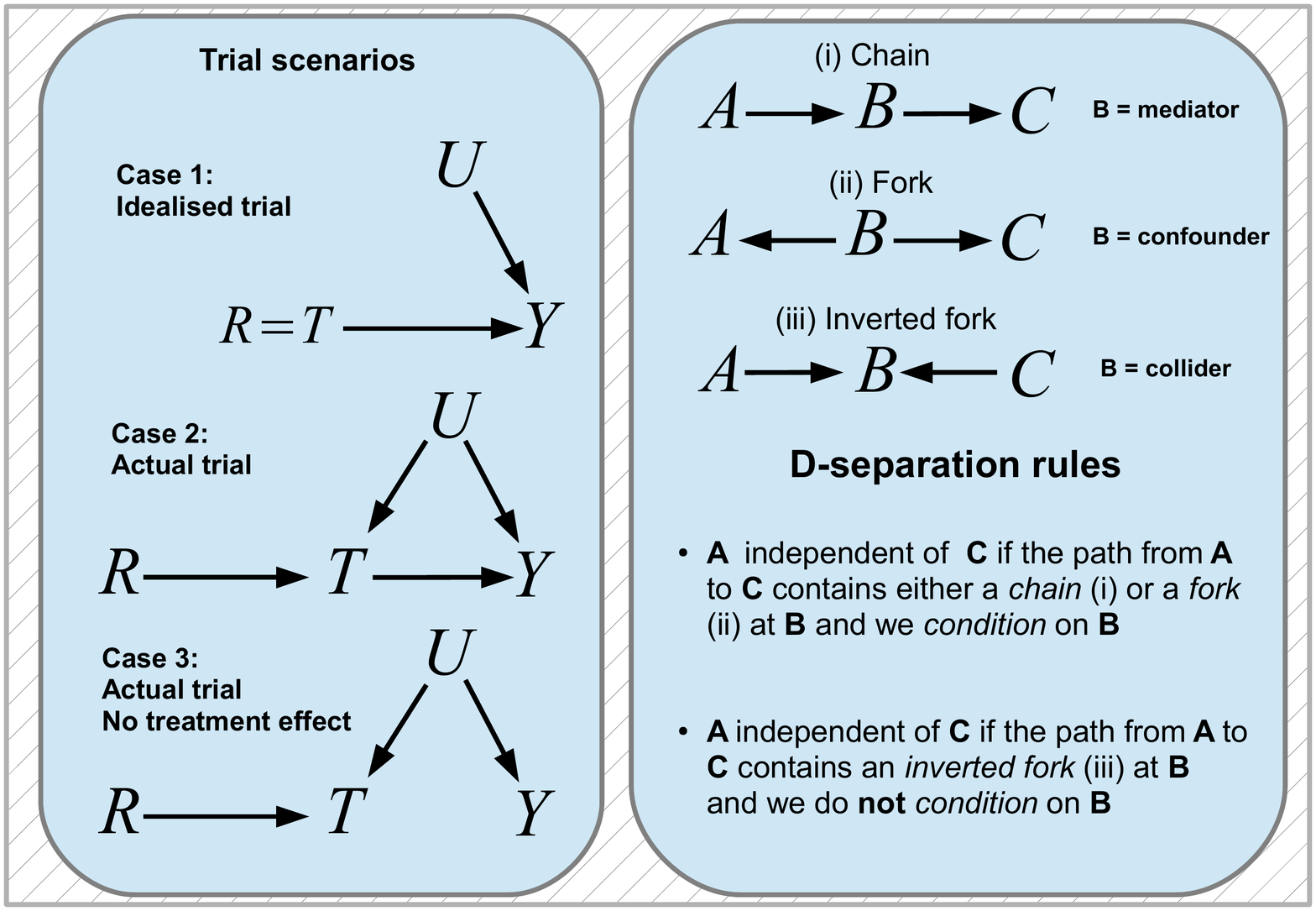}
\includegraphics[bb = 45 69 809 560,width=0.85\textwidth,clip]{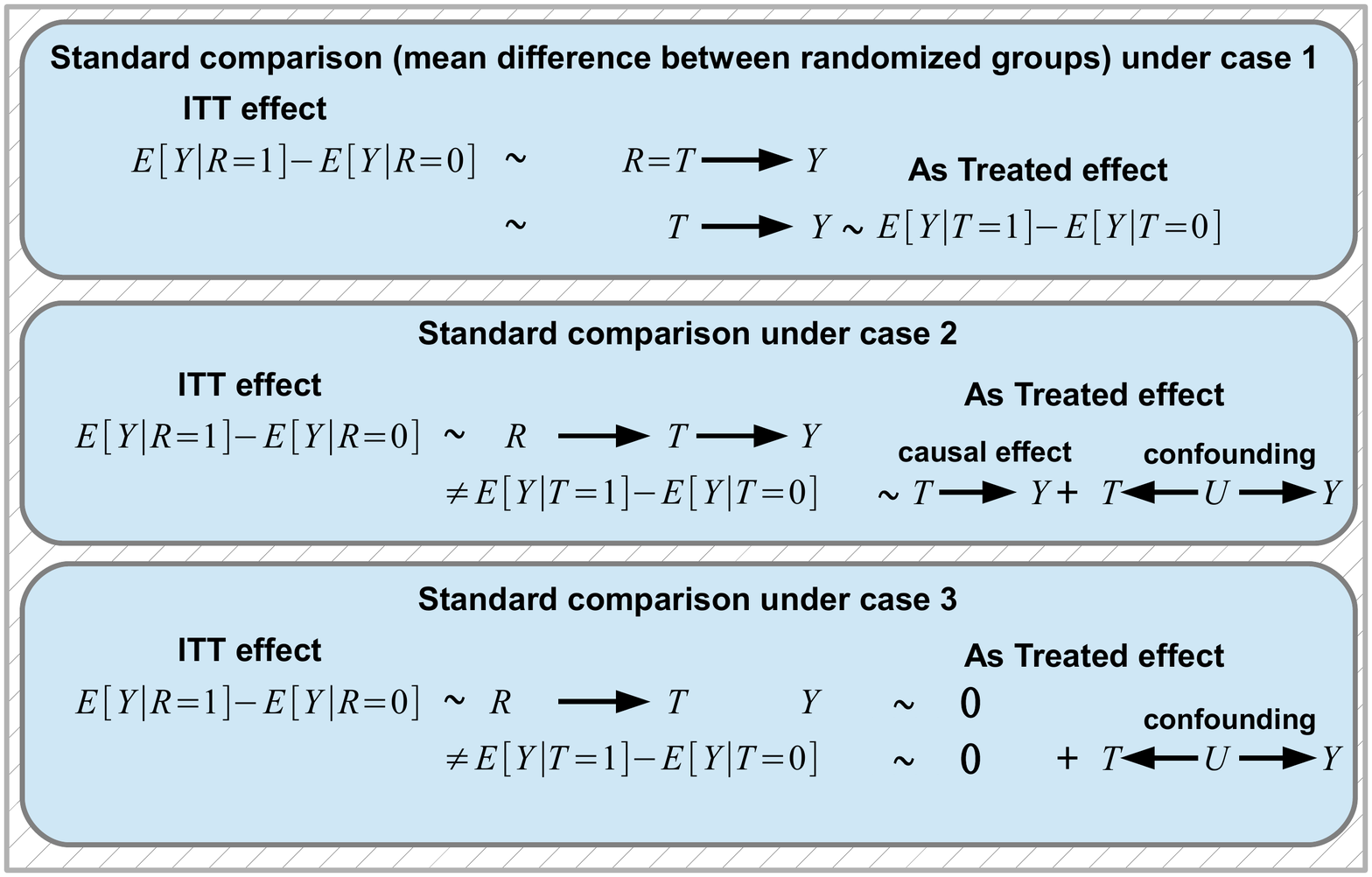}
\caption{{\it Technical box describing Directed Acyclic Graphs (DAGs), the rules of d-separation, and why an As Treated analyses target different estimands in general}}
\label{fig:DagFigure}
\end{figure}

Consider the interpretation of the trial data in an idealised setting where all patients are both expected to take the treatment assigned to them for the duration of the trial and do so. This is represented by the directed acyclic graph (DAG)\cite{pearl1995} in Figure \ref{fig:DagFigure} (top-left panel) and  marked `Case 1'. DAGs are a widely used tool in epidemiology, but less so in clinical trials.  Here the additional variable $U$ represents all unmeasured factors which could in theory jointly influence the trial outcome and a patient/clinician's decision to take the experimental treatment or not.\\
\\
DAGs contain nodes, which in this case are the random variables $R$, $T$, $U$ and $Y$, and directed edges, such as the arrow that goes from $T$ to $Y$ or from $U$ to $T$. An arrow from $T$ to $Y$ indicates that $T$ causally affects $Y$. The variable $U$ is not observed, but allowing for its potential existence is crucial in order to determine if a particular analysis can, in principle, target a causal effect. This, in turn, is relevant for defining subsequent estimands. On an abstract level the DAG encodes a set of relationships about its constituent variables, and these induce certain statistical dependencies between them. Specifically, any two variables are statistically dependent (or associated) if there is an `open' path or route between them, and independent if all paths between them are `blocked'. The status of a particular path as either open or blocked can be deduced by the application of three simple rules on the DAG, which are referred to as the `d-separation' rules \cite{pearl1995}. This is illustrated in Figure \ref{fig:DagFigure} (top-right panel) for hypothetical variables $A$, $B$ and $C$. In Section 3 we define in more detail what we mean by causal effect (and to whom exactly it applies), but initially suppose the causal effect of interest is represented by the direct arrow from $T$ to $Y$, $T \rightarrow Y$.\\ 
\\
In the idealised trial setting (Case 1) there are no post-randomization changes to treatment, so that  $R$ is identical to $T$ for all patients.  Comparing outcomes across randomized groups (an ITT analysis) is then equivalent to comparing outcomes between those that receive treatment and those that do not (an `As Treated' analysis). Now consider a trial with non-adherence, meaning that not all patients receive the treatment they were randomly assigned to, so that $R$ is not always equal to $T$. In this case variables (e.g $U$) almost certainly exist which simultaneously predict (or confound) the treatment-outcome relationship. That is, they play the role of a `fork', as illustrated in Figure \ref{fig:DagFigure} (Top right panel, case (ii)). This scenario is illustrated by the DAG labelled `Case 2' in Figure \ref{fig:DagFigure}.  The  ITT and As Treated analyses in Case 2 are now seen to target different estimands. An ITT analysis would reflect the causal effect of $R$ on $Y$ mediated by treatment $T$. An As Treated analysis would reflect the causal effect $T$ on $Y$ ($\beta$, say)  plus a bias term,  Cov($T,U$)/Var($T$),  reflecting the association between $T$ and $Y$ due to the confounder $U$. This bias arises because the path $T\leftarrow U \rightarrow Y$ is open. It could be blocked by conditioning on $U$ (rule (ii) in Figure \ref{fig:DagFigure}), but this is not possible when at least some component of $U$ is unobserved.  Case 3 in Figure \ref{fig:DagFigure} is the same as Case 2 except that now there is no causal effect of $T$ on $Y$.  In this scenario an ITT analysis would yield a zero estimate because there is no open path from $R$ to $Y$, whereas an As Treated analysis  would estimate the magnitude of the confounding bias, which could be misinterpreted as a treatment effect.

\subsection{The Instrumental Variable Assumptions}

An ITT analysis is a valid tool for investigating causality (i.e. the presence of treatment effect) in Cases 1-3 because $R$ is an IV, as defined by the following three assumptions:

\begin{itemize}
\item{`Relevance' or IV1: $R$ influences $T$, as reflected here by the causal arrow $R\rightarrow T$. This means that they are associated (not independent);}
\item{`Randomization' or IV2: $R$ is independent of $U$ (as denoted by the lack of an open path from $R$ to $U$);}
\item{`Exclusion Restriction' or IV3: $R$ is independent of $Y$ given $T$ and $U$ (as denoted by the lack of an open path from $R$ to $Y$ conditional on $T$ and $U$).} 
\end{itemize}

Assumption IV1 is referred to as the `Relevance' assumption, which is guaranteed to hold whenever patients who are randomized to the experimental treatment are more likely to take it than patients who are randomized to the control. This is almost always true.  Assumption IV2 is known as the `Randomization' assumption. It justifies why it is not necessary to adjust for patient covariates in a randomized trial analysis to remove bias. Assumption IV3 is sometimes referred to as the `Exclusion Restriction'. If it were possible to adjust for all confounders, then in the absence of any treatment effect (Case 3) IV3 implies that the ITT effect is
\begin{eqnarray}
0 &=& E[Y|R=1,T=1,U=u] - E[Y|R=0,T=0,U=u] \nonumber \\
  &=& E[Y|T=1,U=u] - E[Y|T=0,U=u]. \nonumber 
\end{eqnarray}
The equivalence of this hypothetical confounder adjusted treatment effect with the ITT effect under Case 3 follows because conditioning on $U$ blocks all paths from $R$ to $Y$ not through $T$, but in this case $T$ itself exerts no direct effect on $Y$. Assumptions IV2 and IV3 imply together that randomization can only influence the outcome through treatment.  It is highly plausible in double blind trials, but may be violated in cases where patients become unblinded, and alter their behaviour based on the treatment they know they are receiving. It is important to realise that the Exclusion Restriction is not an immutable property of randomization, its validity is context-specific and depends just as strongly on the intercurrent event one wishes to account for. We will return to this issue in Section 5.\\
\\
Assumptions IV1-IV3 are stated in a way that naturally complements the formulation of the causal problem using DAGs. However, they are not unique, and many different formulations exist. For example, in Angrist et al \cite{angrist1996}, the IV assumptions are first described within the context of a two-level structural model for the treatment given the IV and the outcome given the treatment. IV1 is equivalent to the statement in \cite{angrist1996} that $Cov(R,T)\neq$ 0. IV2 and IV3 are equivalent to the statements in \cite{angrist1996} that the $E(R_{i}\epsilon_{i})$ = $E(R_{i}\nu_{i})$ = 0, where $\epsilon_{i}$ and $\nu_{i}$ are `disturbances' in the treatment and outcome model respectively.

\subsection{Potential outcomes}

We now introduce the potential outcomes notation which will be used to define estimands of interest. Let $T_{i}(r=1)$ = $T_{i}(1)$ and $T_{i}(r=0)$ = $T_{i}(0)$  denote the potential treatment received random variable for an individual patient $i$, if they were assigned to treatment or control, respectively. In the same vein let  $Y_{i}(r=1)$ and $Y_{i}(r=0)$  denote their potential outcome under assignment to either treatment. Finally, let $Y_{i}(r=1;t=0)$ = $Y_{i}(1;0)$, $Y_{i}(1; 1)$, $Y_{i}(0; 0)$ and $Y_{i}(0; 1)$ represent the four potential outcomes for patient $i$ when setting randomised treatment and actual treatment to all possible joint values. Only one realisation of each of $T_{i}(r)$, $Y_{i}(r)$ and $Y_{i}(r,t)$ is observable for each patient. For example, if a patient is randomized to the experimental treatment and takes it, we assume that we have observed $T_{i}(r=1)$, $Y_{i}(r=1)$ and $Y_{i}(r=1,t=1)$ for that patient. \\
\\
The Exclusion Restriction is often loosely defined as the statement that randomization only affects the outcome through the treatment. Following Hernan and Robins \cite{hernan2020}, the Exclusion Restriction is defined more formally via the statement that potential outcomes do not depend on $R$. That is:

\begin{equation}
Y_{i}(r=0,t)=Y_{i}(r=1,t). \nonumber 
\end{equation}
This implies that a person's potential outcome given treatment at level $t$ is independent of randomization, so that both
\begin{eqnarray}
Y_{i}(r=0,t=1)&=&Y_{i}(r=1,t=1) \label{eq:ER1} \\
&\text{and}&  \nonumber \\
Y_{i}(r=0,t=0)&=&Y_{i}(r=1,t=0). \label{eq:ER2}
\end{eqnarray}
Under the `full' Exclusion Restriction, we can simplify the notation by writing $Y_{i}(0; 1)$ = $Y_{i}(1; 1)$ =$Y_{i}(t=1)$ and $Y_{i}(0; 0)$ = $Y_{i}(1; 0)$ = $Y_{i}(t=0)$. We will subsequently explore instances where the Exclusion Restriction is fully satisfied, instances where it is `weakly violated' (so that  (\ref{eq:ER2}) holds but (\ref{eq:ER1}) does not), and instances where both (\ref{eq:ER1}) and  (\ref{eq:ER2}) are violated.

\section{Etimands and Estimation}
\label{s:est}

\subsection{Defining and estimating the Treatment Policy estimand}

The Treatment Policy estimand for an individual is the difference between their potential outcomes under assignment to the treatment and control, regardless of the value of treatment received:

\[
Y_{i}(r=1) -  Y_{i}(r=0)
\]

\noindent Since this, or any individual level causal effect is unobservable, we instead define the estimand as a mean difference in observed potential outcomes value across randomized groups: 

\begin{equation}
\text{Treatment Policy estimand:}=E[Y_{i}(r=1)] -  E[Y_{i}(r=0)], \label{eq:policy} 
\end{equation}
It can be viewed as conceptually equivalent to the ITT estimand.  Although we may believe that randomization satisfies the IV assumptions with respect to treatment, we strictly only need the second IV assumption, that $R$ is independent of any confounders of treatment and outcome in order to consistently estimate this quantity via a simple comparison of mean outcomes across randomized groups (in the absence of missing outcome data). 

\subsection{Defining and estimating a Principal Stratum estimand}\label{sec:princStrat}

A Principal Stratum estimand can generally be defined as the treatment effect within the subgroup of participants for whom the intercurrent event would (or would not) occur in the time-frame of the trial under assignment to one or more treatments. This means that several possible Principal Stratum estimands can usually be defined in a given context. We start by considering the canonical example in epidemiology (for example, see \cite{greenland2000,greenland2008,bellamy2007}) where treatment non-compliance is the intercurrent event, and assume we are interested in the principal stratum of patients who would take the treatment if and only if they are randomized to do so. That is, those for whom $T_{i}(r=1)=1$ and $T_{i}(r=0)=0$. This group are generally referred to as `Compliers' ($c$) within a conceptual framework termed Principal Stratification \cite{angrist1996,frangakis2002}, which also defines three additional compliance classes: 

\begin{itemize}
\item{Always Takers ($at$): individuals who would always take the treatment irrespective of treatment assignment, so that $T_{i}(1)=T_{i}(0)=1$;}
\item{Never Takers ($nt$): individuals who would never take the treatment irrespective of treatment assignment, so that $T_{i}(1)=T_{i}(0)=0$;}
\item{Defiers ($d$): individuals who would always dis-respect randomization by taking the treatment not assigned to them, so that  $T_{i}(1)=0$ and $T_{i}(0)=1$.}
\end{itemize}

This framework presupposes that each person is a member of only one compliance class such that membership in a class is not random, but a fixed attribute of every individual. This makes subsequent development more straightforward but can be relaxed \cite{small2017}. The effect of treatment in the Principal Stratum of Compliers, known as the Complier Average Causal Effect (CACE) \cite{frangakis2002} can be expressed as:

\begin{equation}
\text{Principal Stratum estimand:} =   E(Y_{i}(r=1)-Y_{i}(r=0) | T_{i}(1)=1, T_{i}(0)=0)  \label{eq:PS}
\end{equation}

\noindent As a first step to identifying the CACE, we write the expected outcome under assignment $r=1$ or $r=0$ as a weighted average across all compliance classes:

\begin{eqnarray}
E[Y_{i}(r=1)] &=& E_{1c}\pi_{c} + E_{1d}\pi_{d} + E_{1at}\pi_{at} +  E_{1nt}\pi_{nt}, \nonumber \\
E[Y_{i}(r=0)] &=& E_{0c}\pi_{c} + E_{0d}\pi_{d} + E_{0at}\pi_{at} +  E_{0nt}\pi_{nt}. \nonumber 
\end{eqnarray}

\begin{figure}[!h]
    \centering
\includegraphics[bb = 30 80 811 564, width=0.8\textwidth,clip]{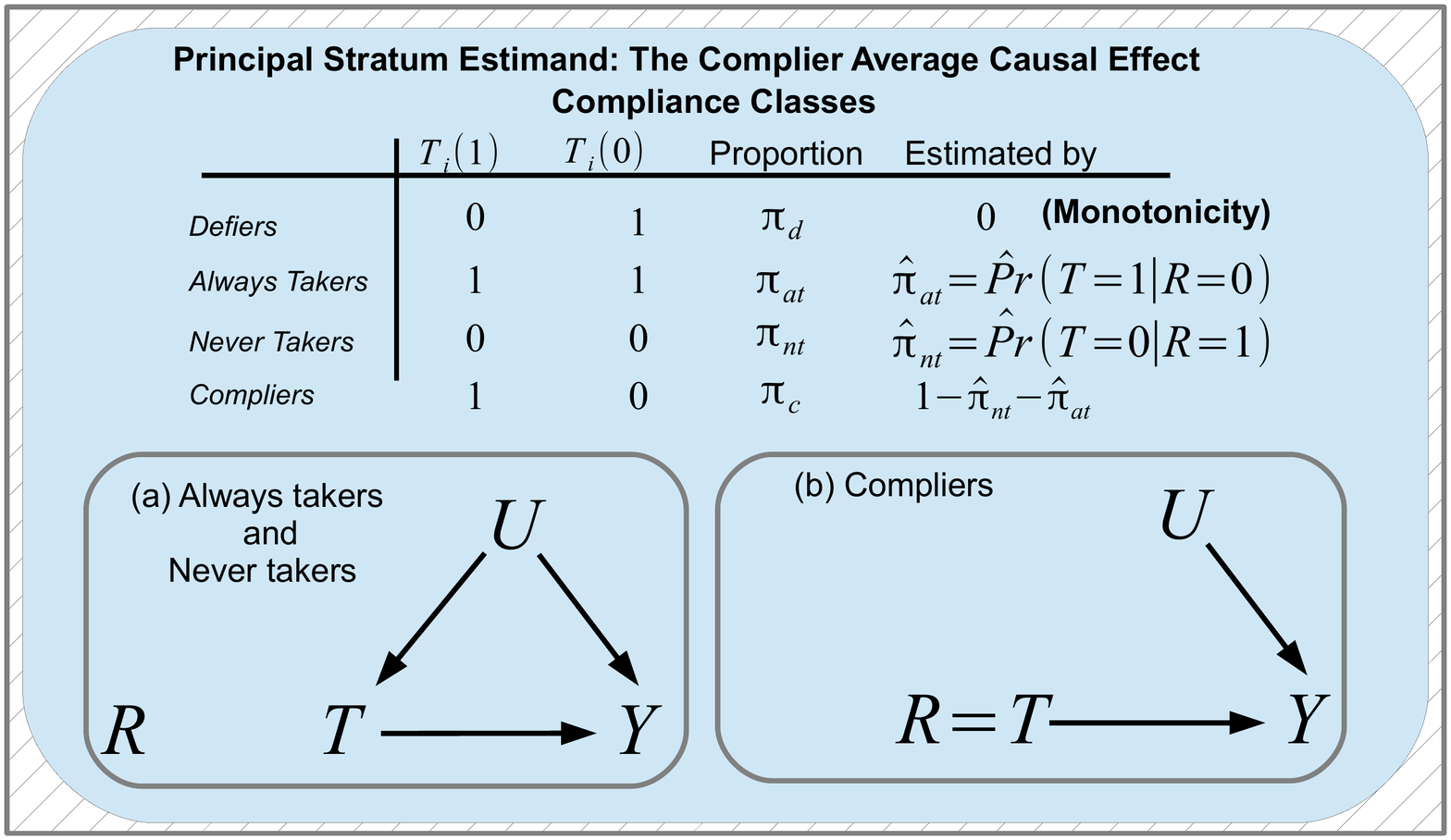}
\caption{{\it An Explanation of Principal Stratification.}}
\label{fig:PrincipalSDags}
\end{figure}

\noindent Here, $E_{j*}$ represents the expected potential outcome under assignment $r=j$ for compliance class $* = (c, at, nt, d)$ and the $\pi$ terms represent their true proportions in the population. Under assumption IV2, these proportions are independent of (or common across) randomized groups. Furthermore, randomization does not affect the outcome for Always and Never Takers, so that $E_{1at}$ =$E_{0at}$ = $E_{at}$ and $E_{1nt}$ =$E_{0nt}$ = $E_{nt}$. Taking the difference of the two expected outcomes removes their contribution completely to leave

\begin{equation}
E[Y_{i}(r=1)]- E[Y_{i}(r=0)] =  CACE\pi_{c} - DACE\pi_{d}, \label{eq:DACE1}
\end{equation}

\noindent where CACE =$E_{1c} - E_{0c}$  and the treatment effect in Defiers (Defier Average Causal Effect, DACE) equals $E_{0d} - E_{1d}$.   In order to identify the CACE, it is generally assumed that Defiers do not exist so that $\pi_{d}=0$. This is referred to as the 'Monotonicity' assumption. Alternative assumptions, such as `Principal Ignorability' (see \cite{feller2017} for an overview)  could instead be invoked but will not be considered further in this paper. The impact of assuming Monotonicity is three-fold: firstly, equation (\ref{eq:DACE1}) reduces to the complier fraction times the CACE; secondly, since they sum to 1, $\pi_{c} = 1 - \pi_{at}  - \pi_{nt}$; thirdly, $\pi_{at}$ can be estimated by the proportion of patients who are randomized to control but take the treatment and $\pi_{nt}$ can be estimated by the proportion who are randomized to treatment but take the control. This means the complier fraction can be estimated as
\begin{eqnarray}
\hat{\pi}_{c} &=& 1 -  \hat{Pr}(T=1|R=0) - \hat{Pr}(T=0|R=1) \nonumber \\
                    &=&       \hat{Pr}(T=1|R=1) - \hat{Pr}(T=1|R=0), \nonumber
\end{eqnarray}
and the CACE can be estimated as the ratio of the Treatment Policy and complier fraction estimates. The rationale for this procedure is further illustrated in Figure \ref{fig:PrincipalSDags}. We note that under the Principal Stratification framework, although the Policy Estimand in the Always Takers and Never Takers is zero, the treatment effect in these groups is left undefined.

\subsection{Defining and estimating the Hypothetical estimand}

The E9 Addendum defines the Hypothetical estimand strategy as defining a contrast between treatment and control in a scenario where the intercurrent event is set to specific hypothetical levels. When the intercurrent event relates directly to treatment, we could ask what the difference in patient outcomes would have been if all patients had received the treatment compared to if none had received the treatment. This suggests the following Hypothetical estimand:

\begin{equation}
\text{Hypothetical estimand:} =E[Y_{i}(t=1) -  Y_{i}(t=0)]\label{eq:SMM2}  
\end{equation}

\begin{figure}[htbp]
    \centering
\includegraphics[width=0.85\textwidth,clip]{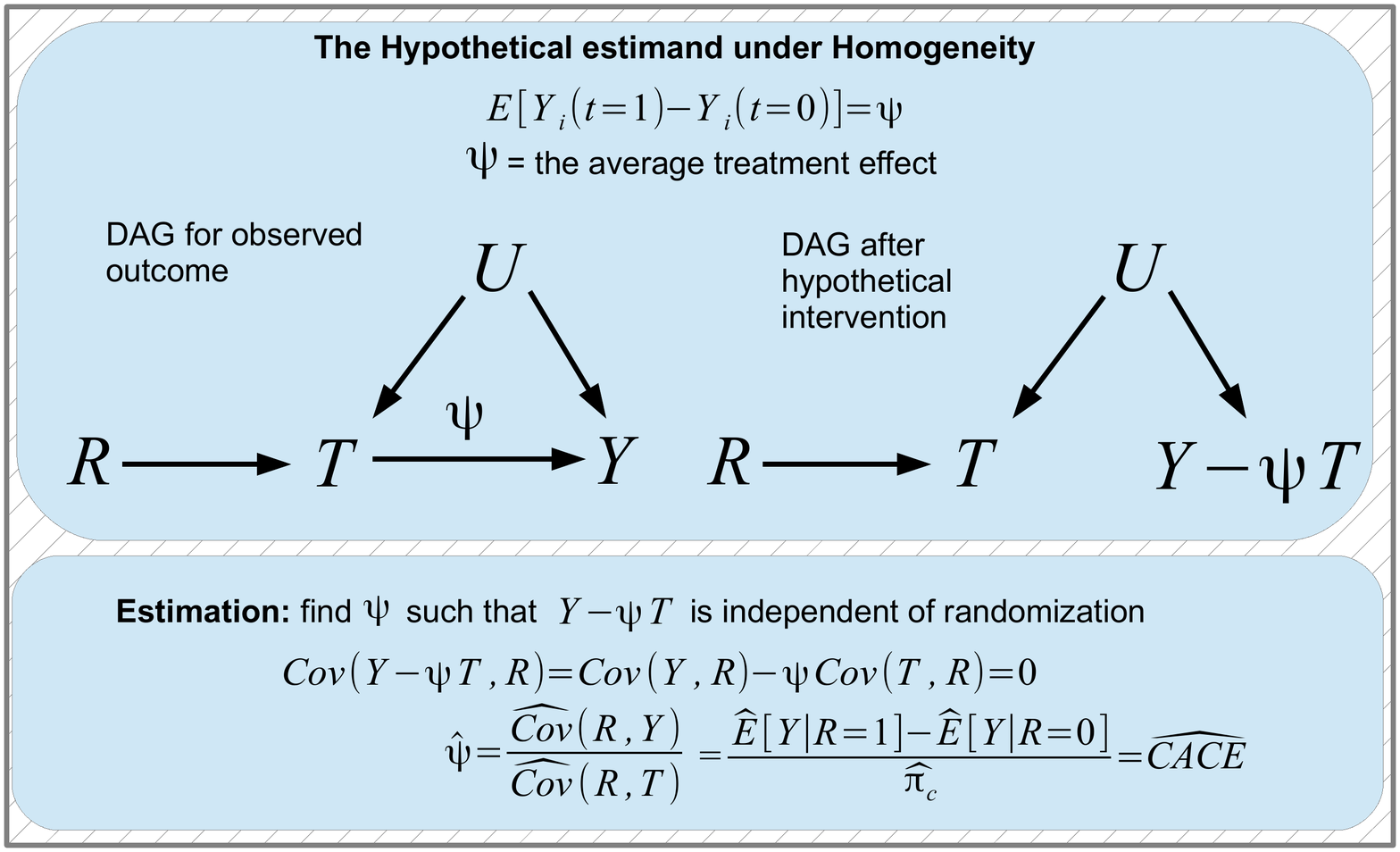}
\caption{{\it Defining and estimating the Hypothetical estimand.}}
\label{fig:SMMs}
\end{figure}

Unlike Principal Stratification, the Hypothetical estimand (\ref{eq:SMM2}) describes an effect in the entire trial population rather than a particular sub-group. This estimand, denoted by $\psi$, can be easily identified if the causal effect of treatment is truly constant across all individuals. This is the simplest but most stringent statement of the `Homogeneity' assumption. We note in passing that this now implies that the treatment effect in Always-Takers and in Never-Takers is not undefined, as it was assumed in section \ref{sec:princStrat}, but rather that compliance class makes no difference to the treatment effect. Hence, the counterfactual setting for the hypothetical estimand differs from that for the principal stratum estimand: It assumes that compliance could be enforced. In Section 5.2 we will introduce some alternative, less stringent definitions of Homogeneity, and how it can be formally tested within an extended regression model.\\
\\
Under Homogeneity, and for a continuous outcome, $\psi$ can be estimated by finding the value of the `treatment-free outcome' that is independent of randomization (or equal across groups) for the trial data (see Figure \ref{fig:SMMs} for details). Note that the observed and treatment-free outcomes are only different for the patients who were actually treated. For a binary outcome the procedure is the same, except $\psi$ would quantify the shift in probability of response under treatment compared to no treatment. In both cases it can be easily shown that $\psi$ is  estimated to be the ratio of the sample covariance between $R$ and $Y$ and the sample covariance between $R$ and $T$. This is equivalent to classical formulation of the IV estimate given by Durbin \cite{durbin1954}.  The DAG intuition for this estimation procedure is illustrated in Figure \ref{fig:SMMs}. Subtracting the treatment effect $\psi$ from $Y$ removes the arrow $T\rightarrow Y$. This means that there is no open path between $R$ and $Y$, hence their independence. 

\subsection{Equivalence of estimates in practice}

\begin{figure}[hbtp]
    \centering
\includegraphics[bb = 10 10 850 320, width=0.9\textwidth,clip]{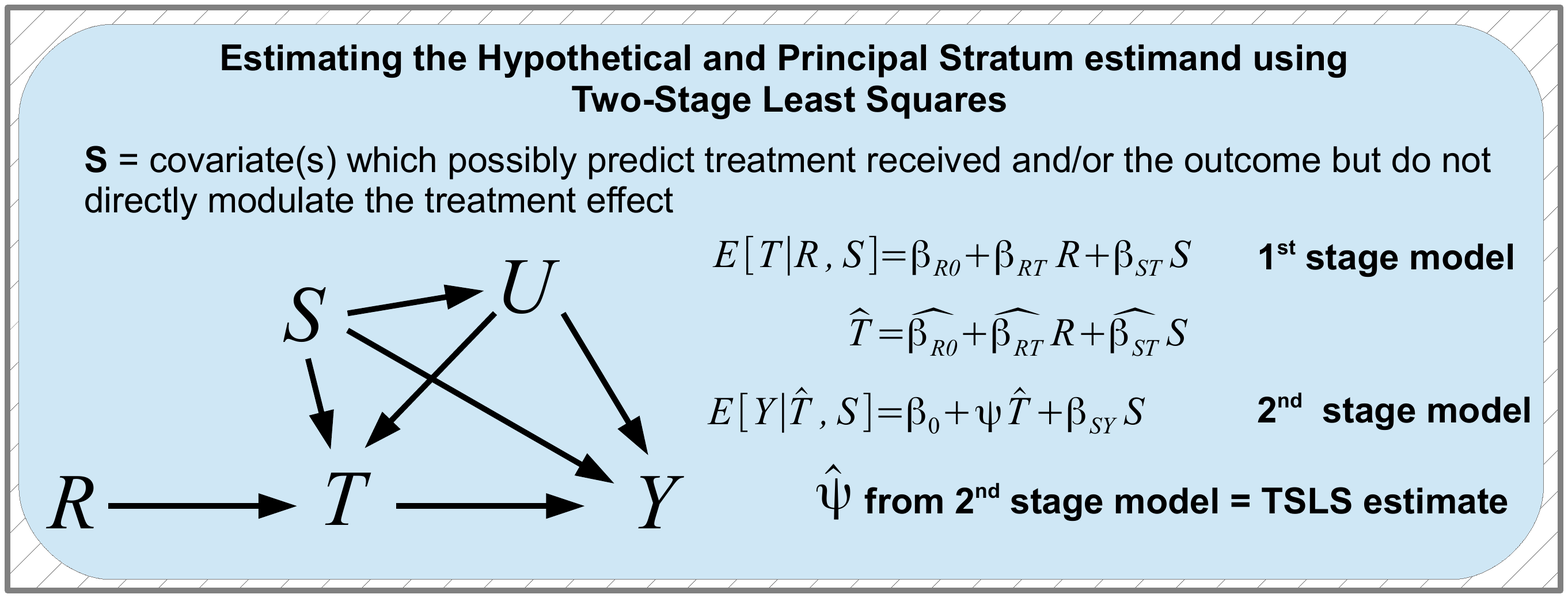}
\caption{{\it Two-Stage Least Squares estimation for estimating the Hypothetical estimand under Homogeneity or the Principal Stratum estimand under Monotonicity}}
\label{fig:TSLS}
\end{figure}

When quantifying the estimand as a mean or risk difference, Hypothetical and CACE estimates defined in Sections 4.2 and 4.3 are identical. In addition to the three IV assumptions, if Homogeneity holds then the this single estimate is consistent for the Hypothetical estimand and the Monotonicity assumption holds then it is consistent for the CACE. If both assumptions hold as well as the IV assumptions, it is a consistent estimate of both, and in this special case the two estimands are identical to each other. The equivalence of the Hypothetical and CACE estimates extends to the setting where the causal estimand is expressed as a risk ratio, but this does not hold when the estimand is expressed as an odds ratio. For further details see \cite{windmeijer2010}.\\
\\
The most straightforward approach for obtaining the common IV estimate is to use Two Stage Least Squares (TSLS), see Figure \ref{fig:TSLS}. TSLS is enacted by firstly regressing treatment received, $T$, on randomization, $R$, using a linear model, to give a predicted value $\hat{T}$. This is identical to the estimated Complier fraction $\hat{\pi}_{c}$. The outcome $Y$ is then regressed on $\hat{T}$, again using a linear model, and its resulting regression coefficient is taken as the TSLS estimate. The rationale for TSLS is that, whilst the observed value of $T$ and $Y$ are confounded, $\hat{T}$ and $Y$ are not. This follows from assumptions IV2 and IV3. Baseline covariates (denoted by $S$ in Figure \ref{fig:TSLS})  can also be easily incorporated into the TSLS model as long as the three IV assumptions are satisfied conditional on $S$. If $S$ does not directly modulate the treatment effect, so that including $S$ in the first and second stage models of Figure \ref{fig:TSLS} does not alter the parameter $\psi$, covariate adjustment can increase the precision of the causal estimate if they help to predict $T$ (given $R$) or $Y$ (given $\hat{T}$), whilst leaving the interpretation of the causal estimate unchanged.   

\section{Assessing Monotonicity and Homogeneity}
\label{s:ass}

\subsection{Relaxing Monotonicity for the Principal Stratum estimand}
 
\begin{figure}[htbp]
    \centering
\includegraphics[bb = 30 100 820 565, width=0.95\textwidth,clip]{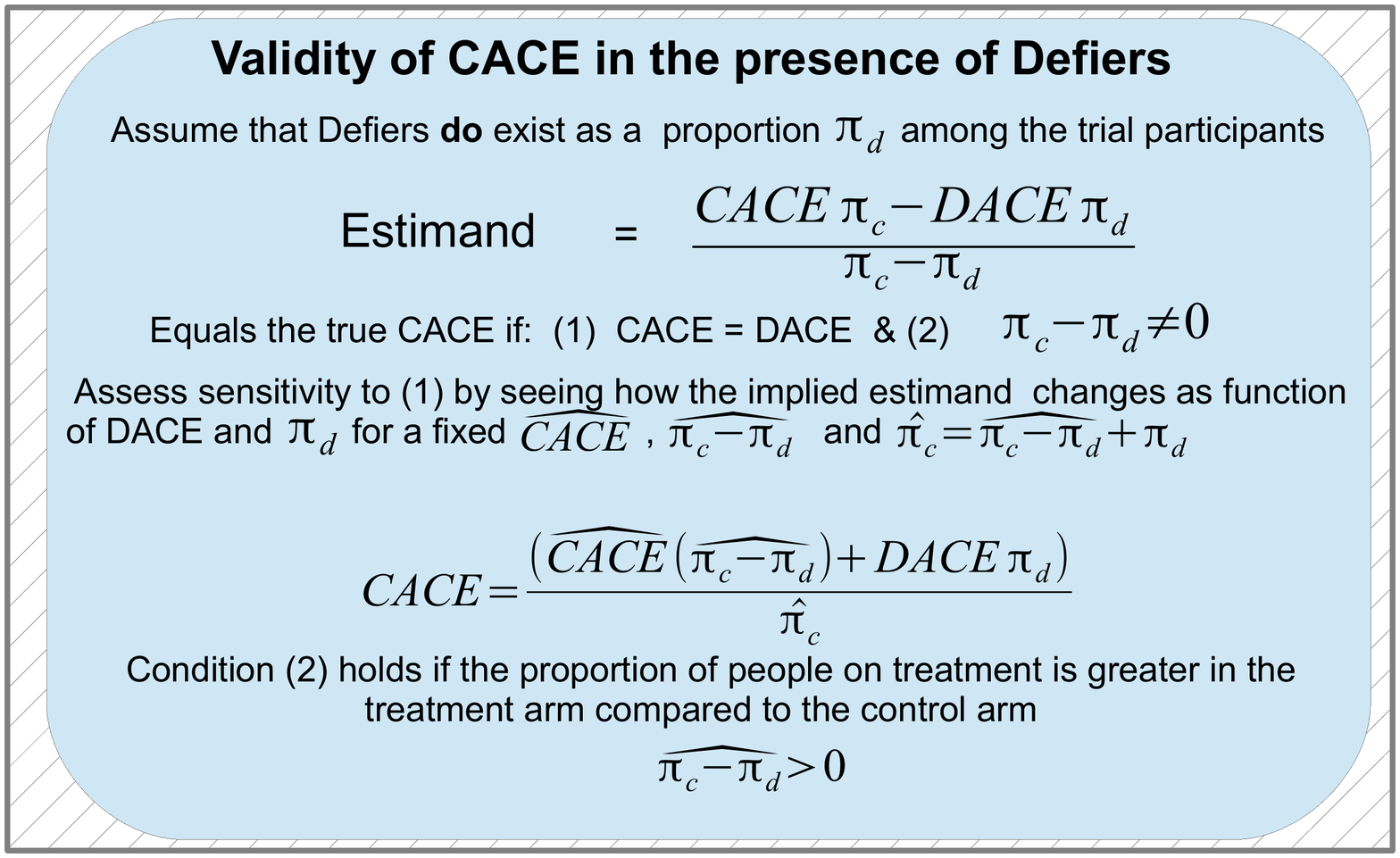}
\caption{{\it Clarifying what is identified by the regular IV estimator in the presence of Defiers}}
\label{fig:Defiers}
\end{figure}

Assessing the plausibility of the  Monotonicity assumption is key to determining whether the CACE can be accurately estimated. In specific circumstances it is  possible to simply rule out the existence of Defiers altogether. For example, suppose that the intercurrent event is defined as receiving treatment when not randomized to receive it, but no one in the control group has the opportunity to receive the treatment at all. In this case we would assert that the proportion of both Always Takers and Defiers in the trial is zero, and consider the trial to be a mixture of Compliers and Never Takers only. Although one could argue that the make up of the Never Taker stratum would be different in this setting compared to a trial where the control group could theoretically access the treatment, we would still need to apply the framework of Principal Stratification as before to recover the CACE. \\ 
\\
When the Monotonicity assumption is violated, the CACE estimate (as given in Figure 3)  instead targets  

\begin{equation}
\frac{CACE\pi_{c} - DACE\pi_{d}}{\pi_{c}-\pi_{d}}. \nonumber 
\end{equation}
 if (i) the DACE is the same as the CACE and (ii) the proportion of Defiers not equal to the proportion of Compliers (so that $\pi_{c}-\pi_{d}\neq 0$), then the CACE can still be consistently estimated without the assumption of Monotonicity. Assumption (i) is essentially saying that the average effect is the same among the Compliers and Defiers.  Assumption (ii) can be verified whenever the probability of receiving treatment across the trial is greater when assigned to it than when not (if this probability is equal across randomized groups then randomization fails assumption IV1 too). In order to allow for Defiers and for violation of (i), a sensitivity analysis could be performed to gauge the impact different values of the DACE and Defier fraction would have on the implied CACE estimand for a given value of the CACE and complier fraction estimators, as shown in Figure \ref{fig:Defiers}. 


\subsection{Relaxing Homogeneity and the Exclusion Restriction for the Hypothetical estimand.}

The Homogeneity assumption facilitates estimation of the Hypothetical estimand, which can be interpreted as the effect of treatment within the entire population. Homogeneity is clearly satisfied if the causal effect is the same for everyone. Although this sufficient condition is implausible, it is not strictly necessary. Several weaker but sufficient definitions of Homogeneity are provided in \cite{hernan2020}. For example, when the treatment (or intercurrent event) is binary, the Hypothetical estimand can be identified if the average treatment effect is constant across randomized groups at each level of the treatment:
{\small \begin{equation}
E[Y(T=1)-Y(T=0)|R=1,T=t] = E[Y(T=1)-Y(T=0)|R=0,T=t] \label{eq:Homogeneity}
\end{equation}}
In order to relax this assumption further, we could instead assume that

{\small \begin{eqnarray}
E[Y(T=1)-Y(T=0)|R=1,T=1] = E[Y(T=1)-Y(T=0)|T(1)=1]  &=& 
\psi_{t} \nonumber \\
E[Y(T=1)-Y(T=0)|R=0,T=1] = E[Y(T=1)-Y(T=0)|T(0)=1] &=& \psi_{at} \nonumber
\end{eqnarray}}

for two possibly distinct parameters $\psi_{t}$ and $\psi_{at}$. Another way of expressing this as the expected difference in potential outcomes for individual $i$ given allocation to $R=r,T=t$ and allocation to $R=r',T=t'$:

\[\
E[Y_{i}(r,t)-Y_{i}(r',t')] = \psi_{t}(tr-t'r') +  \psi_{at}\{t(1-r) - t'(1-r')\}
\]
When the homogeneity assumption is violated $\psi_{t} \neq \psi_{at}$, this model violates the Exclusion Restriction among the treated population, since $E[Y_{i}(1,1)-Y_{i}(0,1)]$ = $\psi_{t}-\psi_{at}$, so that the equality in Equation (\ref{eq:ER1}) does not hold. For ease of interpretation we will assume that Monotonicity holds, so that $\psi_{at}$ pertains to the Always Takers. The model parameter $\psi_{t}$ then represents the Hypothetical estimand within the union of the Compliers and Always Takers. As such, $\psi_{t}$ can then be viewed as a weighted average of $\psi_{at}$ and the hypothetical estimand in the Compliers, $\psi_{c}$.

\[\
\psi_{t} = \frac{\psi_{c}\pi_{c} + \psi_{at}\pi_{at}}{\pi_{c}+\pi_{at}}
\]

The Homogeneity assumption is satisfied if $\psi_{t}$ = $\psi_{at}$, but is violated otherwise. Interestingly, even when the Homogeneity assumption is violated, the standard IV estimate does not target $\psi_{t}$ or $\psi_{at}$, but rather $\psi_{c}$ since: 
\begin{eqnarray}
\frac{Cov(R,Y)}{Cov(R,T)}&=& \frac{\psi_{t}Pr(T=1|R=1)-\psi_{at}Pr(T=1|R=0)}{\pi_{c}} \nonumber \\
&=& \frac{\psi_{t}(\pi_{c}+\pi_{at})-\psi_{at}\pi_{at}}{\pi_{c}} \nonumber \\
&=& \psi_{c} \nonumber
\end{eqnarray}

\begin{figure}[htbp]
    \centering
\includegraphics[bb = 10 200 850 580, width=0.8\textwidth,clip]{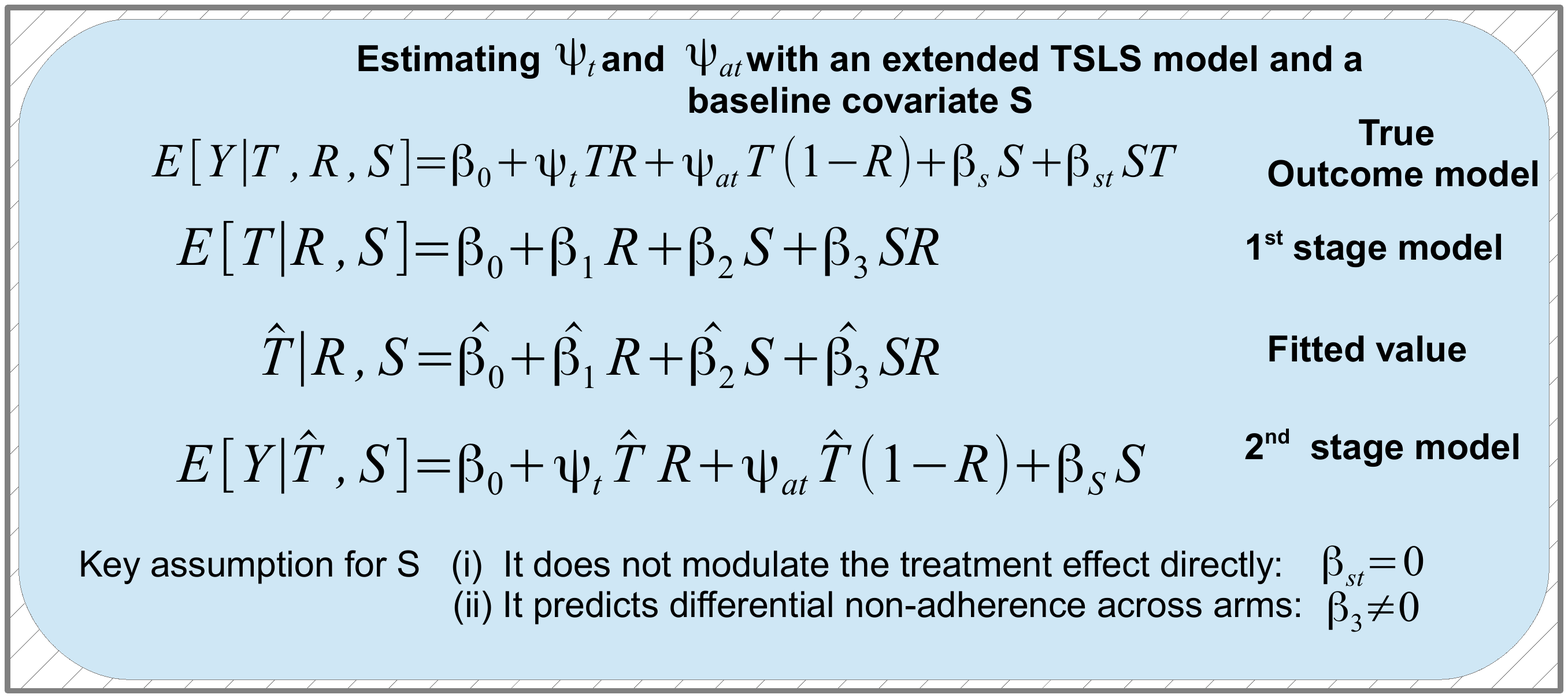}
\caption{{\it Implementation of the two parameter model with an extended TSLS framework}}
\label{fig:relaxHomogeneity2}
\end{figure}

Although this is reassuring on one level, as a practitioner one may instead prefer to report $\hat{\psi}_{t}$ along with the difference $\hat{\psi}_{t}-\hat{\psi}_{at}$ as a sensitivity analysis to the primary Hypothetical estimand analysis. The advantage of $\psi_{t}$ over $\psi_{c}$ is that the former reflects an effect for an observable subset of patients, whereas the latter does not. Figure \ref{fig:relaxHomogeneity2} describes an extended TSLS model to achieve this aim, by estimating $\psi_{t}$ and $\psi_{at}$ separately. In order to do this it requires a baseline covariate, $S$, which satisfies two properties: Firstly, it does not directly modulate the effect of treatment, as indicated by a zero $T\times S$ interaction in the true outcome model (a main effect for $S$ {\it is} allowed). Secondly, it modulates the strength of randomization as an IV across the treatment groups, as indicated by a non-zero $R\times S$ interaction in the model for $T$ given $R$ and $S$. Note that if $S$ were unobserved, this model would itself imply violation of condition H2. For further examples of this approach applied to IV analyses applied in clinical trials and epidemiology see \cite{small2012} and \cite{spiller2019}.\\
\\
In the first stage of the extended TSLS procedure, $T$ is regressed on $R$,  $S$, and $R\times S$. The fitted value from this first stage regression is then regressed on the two-parameter outcome model along with the covariate $S$. The corresponding regression coefficients are then consistent estimates for $\psi_{t}$ and $\psi_{at}$ under the stated assumptions. A formal test for Homogeneity violation could be constructed based on $\hat{\psi}_{t}-\hat{\psi}_{at}$ being significantly different from zero.

\subsection{Fitting TSLS models for continuous, binary and time-to-event outcomes}

When the trial outcome is continuous, the TSLS model in Figure 4 and the extended TSLS model in Figure 6 can be fitted using straightforward linear regression in order to deliver estimates on the mean difference scale. When the exposure and/or outcome is binary, we can replace this with a logistic or probit regression. Taking the extended TSLS model estimate as an example and employing logistic regression, the two level model would be:

\begin{eqnarray}
logit\left\{\pi(T_{i}=1|R_{i},S_{i})\right\} &=& \beta_{0} + \beta_{1}R_{i} + \beta_{2}S_{i} + \beta_{3}R_{i}S_{i}, \nonumber \\
logit\left\{\pi(Y_{i}=1|\hat{T}_{i},S_{i})\right\} &=& \gamma_{Y0} + \psi_{t}\hat{T}_{i}R_{i}+ \psi_{at}\hat{T}_{i}(1-R_{i}) + \gamma_{YS}S_{i}, 
\label{eq:logistic}
\end{eqnarray}

One can then extract the relevant TSLS estimate as an average risk difference, or Average Marginal Effect \cite{gelman2007} via:
\begin{equation}
\frac{1}{n}\sum^{n}_{i=1}\left\{\hat{\pi}(Y_{i}=1|\hat{T}_{i}R_{i}=1,S=s_{i}) - \hat{\pi}(Y_{i}=1|\hat{T}_{i}R_{i}=0,S=s_{i}) \right\}\nonumber
\end{equation}
where $n$ is the sample size and $\hat{\pi}(\hat{T}=t^{*}_{i},S=s_{i})$ is the estimated fitted value for $Y_{i}$ on the probability scale obtained by fitting model (\ref{eq:logistic}). This can easily be calculated using the \verb margins()  package in \verb R. For time-to-event outcomes, we recommend analysing the data using an Aalen additive hazard model in order to yield estimates on the hazard difference scale. Estimating effects as mean differences, risk differences or hazard differences is intended to ensure that they remain constant when marginalised over different sets of unobserved confounders, because they are collapsible measures \cite{tchetgen2015,huitfeldt2019}. This makes it more straightforward to compare estimates from different methods.

\subsection{Simulation example}

\begin{figure}[htbp]
    \centering
\includegraphics[width=0.49\textwidth,clip]{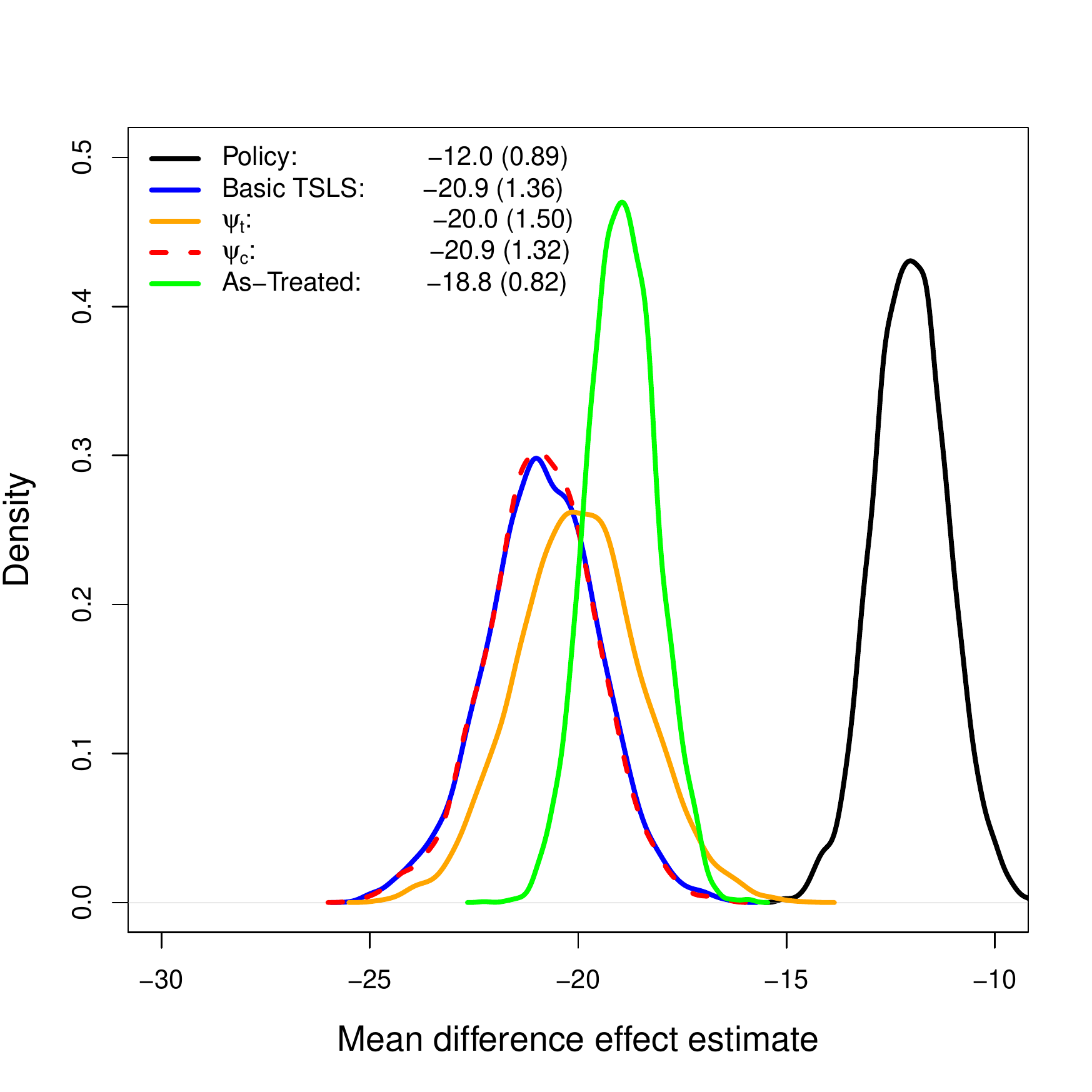}
\includegraphics[width=0.49\textwidth,clip]{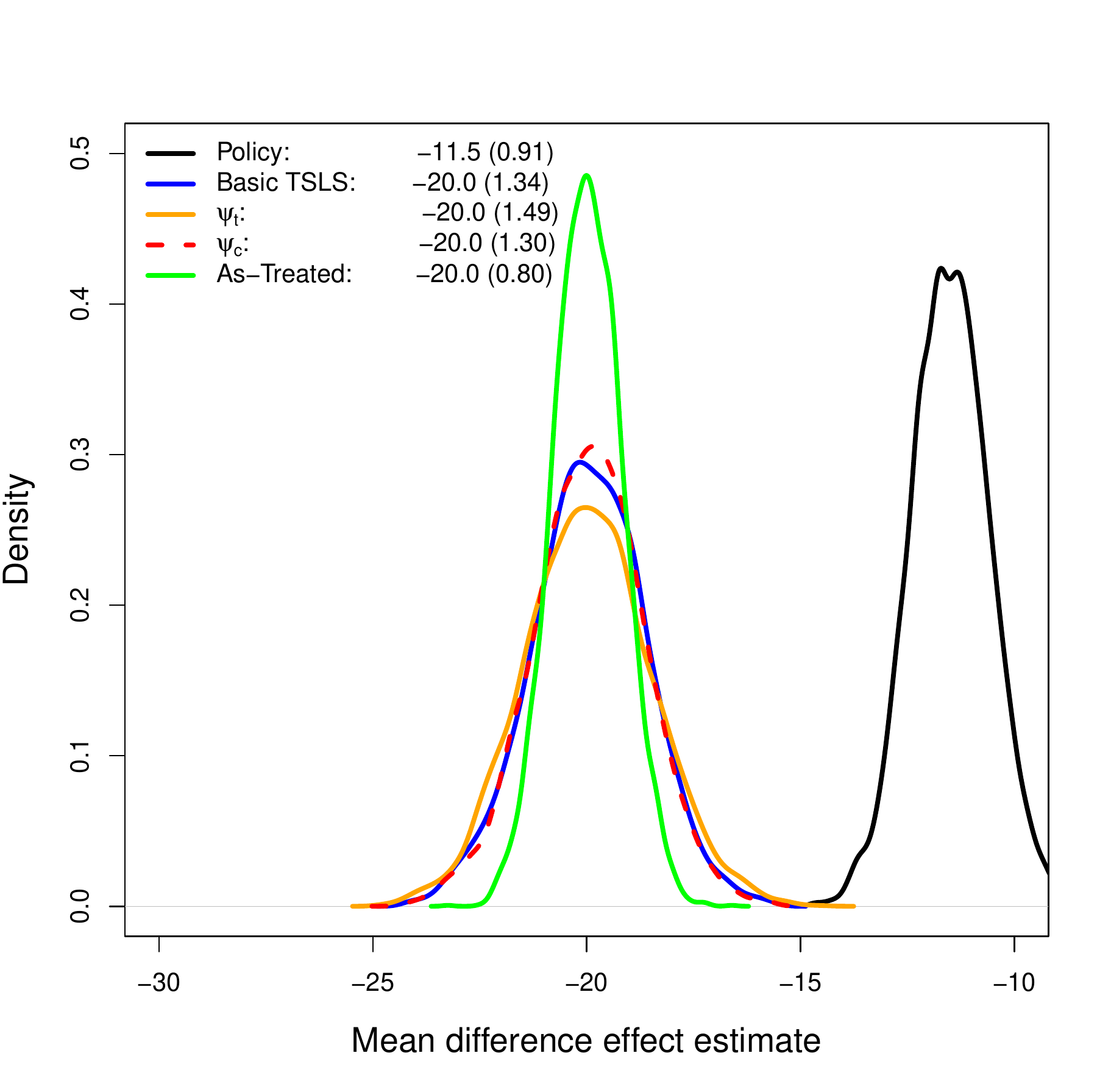}
\caption{{\it Left: Distribution of: the Treatment Policy estimate (black);TSLS estimate (blue);  $\hat{\psi}_{t}$ from an extended TSLS model under a relaxation of the Homogeneity assumption (orange); the implied  parameter estimate for $\hat{\psi}_{c}$ (orange) from the same extended TSLS model (red, dashed); the As Treated estimate (green). Right: Distribution of estimates under Homogeneity and random non-compliance (or no unmeasured confounding).}}
\label{fig:IVreg2}
\end{figure}

In order to elucidate the methods described we simulate data for a placebo controlled trial testing pain relief medication on 1000 participants consistent with the true outcome model in Figure \ref{fig:relaxHomogeneity2}, and thus in violation of the Homogeneity assumption. Further details on the simulation model are given in Appendix A. The trial outcome $Y$ is continuous and is imagined to be a pain severity score between zero and 100, as in \cite{gridley2001}, with a mean of 55 and standard deviation of 15. The mean probability of taking treatment is  65\% in those randomized to receive it, but on average 7\% of patients randomised to control also take the treatment. The true Complier fraction $\pi_{c}$ is therefore estimated to be approximately 58\%. However, the binary covariate $S$ which denotes whether an individual has a previous history of migraine is a strong predictor of treatment. The probability of receiving treatment in the treatment arm among those with a history of migraine, $S$=1, is 97\%, but is only 31\% among those with $S$=0. The average causal effect of treatment for those who are both randomized to the treatment group and take treatment, $\psi_{t}$, equals $-20$ (so that treatment lowers the pain score by 20 points). The average causal effect of treatment for those who are both randomized to the control group and take treatment, is a 10 point reduction ($\psi_{at}$ = -10). This means that the Homogeneity assumption is violated. Using the formula in Figure 6, we can infer that the causal effect of treatment in the Compliers, $\psi_{c}$, is approximately -20.9. Figure \ref{fig:IVreg2} (left) shows density plots across 2000 simulated trials for:
\begin{itemize}
\item{The standard Treatment Policy estimate, as assessed by the mean difference in outcomes across randomised groups;}
\item{The basic TSLS estimate, as fitted in Figure \ref{fig:TSLS} which correctly targets  the  Principal Stratum (CACE) and Hypothetical estimands under Monotonicity and Homogeneity, respectively; and}
\item{The parameter estimates  for $\psi_{t}$ and $\psi_{c}$ obtained from fitting the extended TSLS model under a relaxation of the Homogeneity assumption in Figure \ref{fig:relaxHomogeneity2}.}
\item{The As Treated Estimate, calculated by the mean difference between those who are treated and untreated}
\end{itemize}
Note that although $\psi_{c}$ is not an explicit parameter in the extended TSLS model it can be derived from the estimates for $\psi_{t}$ and $\psi_{at}$.\\
\\
We see the following: The Treatment Policy estimate is the most precise of all the presented estimates, but also the closest to zero. The  basic TSLS estimate has a mean value $-20.9$ as predicted. The extended TSLS model estimate for $\psi_{t}$, allowing for Homogeneity violation, is unbiased for the corresponding Hypothetical estimand among the treated population. Using the extended TSLS model estimates for $\psi_{t}$ and $\psi_{at}$ to derive the implicit CACE, $\psi_{c}$, we see that it agrees perfectly with the standard TSLS estimate, as the theory suggests. 

\subsubsection{When is the `As-Treated' analysis an efficient estimate of the hypothetical estimand?}

In Figure 1 we used DAGs to describe why an As-Treated analysis - that is an analysis comparing outcomes between treated and untreated individuals does not in general give a consistent estimate for the causal effect when there is non-compliance. This motivated the use of the IV approach. However, there is a specific scenario where it not only consistently estimates the causal effect, but is more efficient than an IV analysis: namely when non-compliance is random with respect to the outcome. This is equivalent to the assumption that there is no unmeasured confounding. In that case, the use of all patients in the sample (as in the As-treated estimate) does not cause a bias in the estimates of $\psi_{c}, \psi_{at}, \psi_{nt}$ or $\psi_d$ since these are all the same. At the same time, the use of all patients without any adjustment for potential compliance class differences is of course more efficient statistically than using an estimate which adjusts for non-existing differences between compliance class outcomes (as the TSLS estimate does). Figure \ref{fig:IVreg2} (right) illustrates this. It shows the distribution of  all estimates under the same data generating model as before, except that non-compliance is now random (because the confounder $U$ has been adjusted for) and the Homogeneity assumption is satisfied ($\psi_{at} = \psi_{t}$ = -20). It shows that both the TSLS and As-Treated analyses unbiasedly estimate the hypothetical estimand (-20), but that the standard deviation of the As-Treated estimate (0.82) is approximately $\pi_{c}$ times the standard deviation of the TSLS estimate (1.36).

\section{Clinical trial examples revisited}
\label{s:exrev}

\subsection{Setting 1: adjusting for biomarker response}

\begin{figure}[htbp]
\centering
\includegraphics[bb = 30 30 815 575, width=0.8\textwidth,clip]{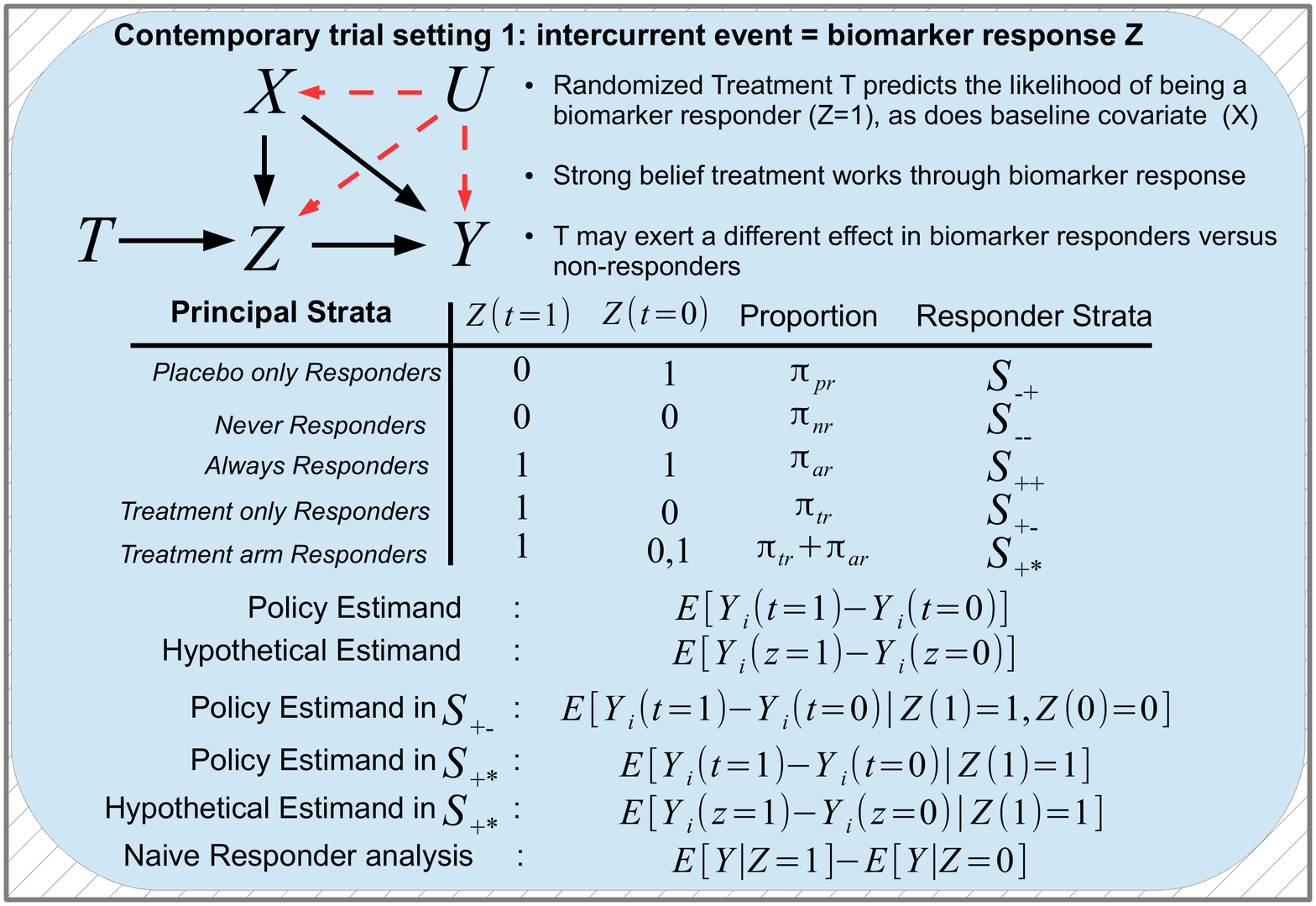}
\includegraphics[ width=0.55\textwidth,clip]{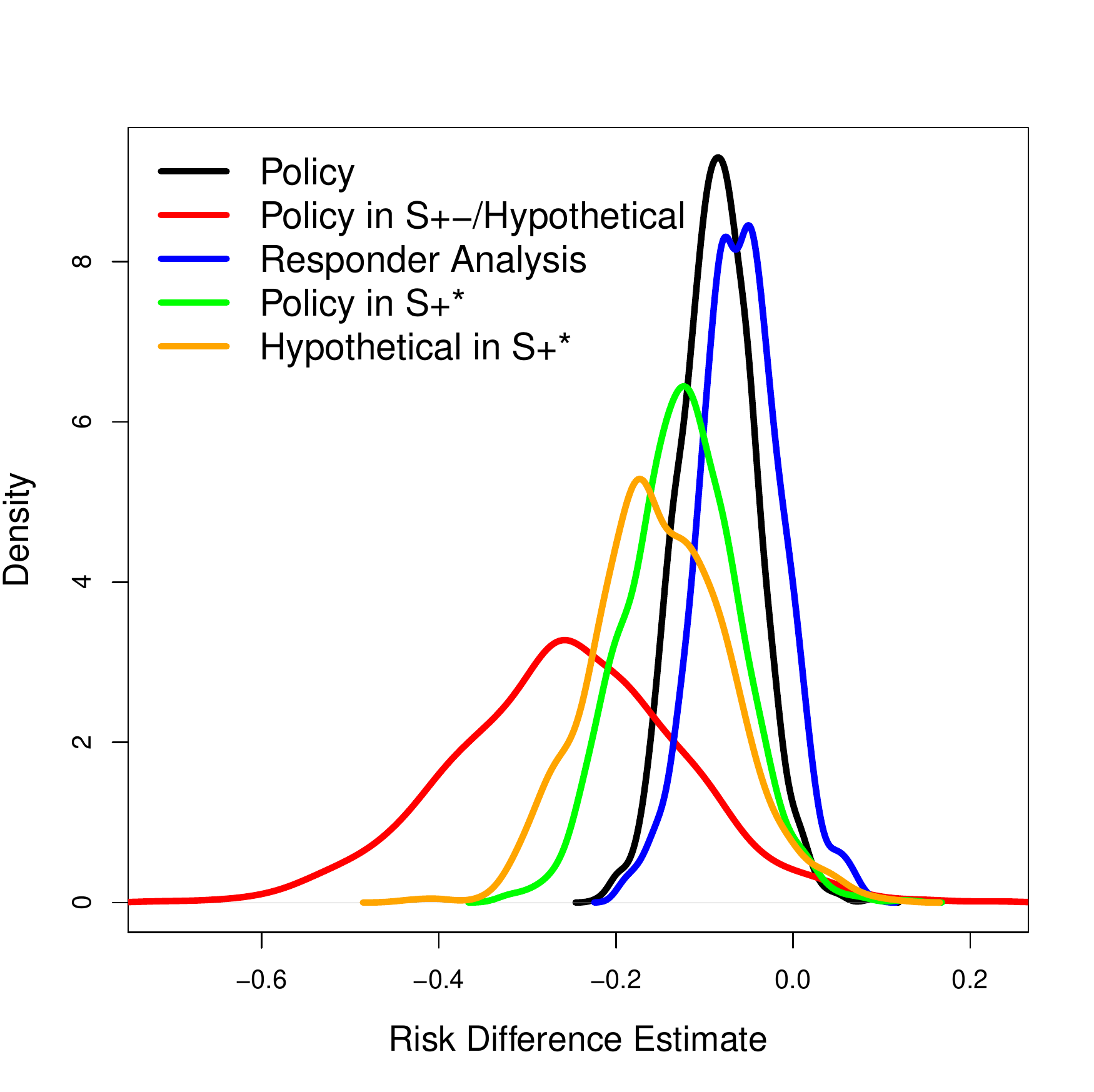}
\caption{{\it Top: An Instrumental Variable formulation contemporary industry trial setting 1: Intercurrent event = biomarker response. Bottom: Distribution of estimates for the simulation study of Section 6.1.2}}
\label{fig:EstimandFigure}
\end{figure}

We now apply the IV methods described in the previous sections to simulated data closely matching the first clinical trial example from Section~\ref{s:ex} to further motivate and clarify the ideas. The DAG in Figure \ref{fig:EstimandFigure} (top) illustrates our assumed trial set up. Firstly, all patients who are randomized to a treatment group take their assigned treatment, meaning that there is no non-compliance in the sense described in the previous section and $R\equiv T$. We assume that treatment $T$, a baseline covariate, $X$, and an unmeasured confounder, $U$ jointly predict whether an individual is deemed to be a biomarker responder ($Z$=1) or a non-responder ($Z$=0)  The binary mortality outcome, $Y$, is assumed to be predicted by $Z$, $X$, $U$ and the effect of $Z$ is possibly modulated by $T$. The variables $X$ and $U$ comprise the set of all confounders of $Z$ and $Y$. Our subsequent analyses will treat biomarker response as the intercurrent event which we believe mediates the effect of treatment on the outcome. Randomized treatment $T$ is a valid IV for $Z$: associated with $Z$ (IV1); independent of $X$ and $U$ (IV2); and only affects $Y$ through $Z$ (IV3).\\
\\
The four traditional Principal Strata considered in the trial are as follows:  Always Responders ($ar$) are those for whom $Z(1)$ = $Z(0)$ =1. Never Responders ($nr$) are those for whom $Z(1)$ = $Z(0)$ = 0. Treatment-only Responders ($tr$) respond if and only if randomized to treatment, so that  $Z(1)$ = 1 and $Z(0)$ = 0. Lastly, Placebo Responders ($pr$) respond if and only if randomized to placebo, so that  $Z(1)$ = 0 and $Z(0) = 1$. In Figure \ref{fig:EstimandFigure} (top) we show how these strata can be expressed using the `S' notation of Qu et. al, which we call `Responder Strata'.   \\

\subsubsection{Trial estimands}

A list of the trial estimands considered for the trial is given in Figure \ref{fig:EstimandFigure} (top). The Policy Estimand is the average difference in potential outcomes under randomization to treatment and control, irrespective of whether the intercurrent event (biomarker response) occurred or not. This estimand is identified as long as randomization was adequately performed.\\
\\
Several Principal Stratum estimands are conceivable. One possibility is the Treatment Policy Estimand in the $S_{+-}$ stratum of patients who would have been a biomarker responder under allocation to treatment, and who would not if allocated to the control. This is directly analogous to the CACE estimand described in Section 4. It can be identified if $T$ is a valid IV and the Monotonicity assumption (no Placebo Responders) holds and estimated using the standard IV formula (Figure 3). Placebo Responders are those whose body would naturally produce the correct biological response without treatment, but who would not do so if given the treatment. The treatment itself would therefore have to disturb the body's natural response in this group. The Hypothetical Estimand is the expected difference in potential outcomes if all patients were forced to be biomarker responders compared to if all patients were forced to be non-responders. It is identified if $T$ is a valid IV and the Homogeneity assumption holds and can again be estimated using the standard IV formula. We are, however, willing to countenance the possibility the Homogeneity assumption is violated and that biomarker response exerts distinct, non-zero effects in treated and untreated individuals. That is, the expected difference in potential outcomes for individual $i$ given $T=t,z=z$ and $T=t',z=z'$ satisfies:
\[\
E[Y_{i}(t,z)-Y_{i}(t',z')] = \psi_{+*}(zt-z't') +  \psi_{++}\{z(1-t)-z'(1-t')\}
\]
Our main quantity of interest in this model, the  parameter $\psi_{+*}$ represents the Hypothetical Estimand in the $S_{+*}$ stratum, which is the union of the Treatment-only Responders  and Always Responders. As shown in the previous section, we can view the parameter $\psi_{+*}$ as a weighted average of these two groups:
\begin{equation}
 \psi_{+*} = \frac{\psi_{+-}\pi_{tr} + \psi_{++}\pi_{ar}}{\pi_{tr}+\pi_{ar}} \nonumber    
\end{equation}
where $\psi_{+-}$ is the treatment effect in Treatment-only Responders. We can consistently estimate $\psi_{+*}$ by fitting a two parameter TSLS model using $T$ and $X$ if $T$ is a valid IV and $X$ differentially predicts biomarker response across groups, without modulating the treatment effect. The parameter $\psi_{++}$ represents the Hypothetical Estimand in Always Responders only. It is not of direct interest in this context, except in the case that the Homogeneity assumption is satisfied and $\psi_{+*}= \psi_{++}$.\\
\\
The final estimand we will consider is the Policy Estimand in the $S_{+*}$ stratum 
\[\
E[Y_{i}(t=1)-Y_{i}(t=0)|Z(1)=1],
\]
which is an alternative principal stratum estimand. B{\"o}rnkamp and Bermann \cite{bornkamp2019} proposed methodology within the Estimand Framework to target this quantity. For identification, they assumed that all confounders of biomarker response and the outcome were known, and they used weighting and standardisation techniques for estimation. Here we show that it can alternatively be estimated using IV methods, by re-writing it as

\begin{equation}
\sum^{1}_{j=0}E[Y_{i}(t=1)-Y_{i}(t=0)|Z(1)=1,Z(0)=j]Pr(Z(0)=j|Z(1)=1). \nonumber
\end{equation}
In this context
\begin{eqnarray}
Pr(Z(0)=0|Z(1)=1) &=& \frac{\pi_{tr}}{\pi_{tr}+\pi_{ar}}, \nonumber \\
Pr(Z(0)=1|Z(1)=1) &=& \frac{\pi_{ar}}{\pi_{tr}+\pi_{ar}}, \nonumber
\end{eqnarray}
are the conditional probabilities of being a Treatment Responder and Always Responder, respectively, given $Z(1)=1$. This Principal Stratum estimand can then be written as
\begin{equation}
\psi_{+*}\frac{\pi_{tr}}{\pi_{tr}+\pi_{ar}} +(\psi_{+*}-\psi_{++})\frac{\pi_{ar}}{\pi_{tr}+\pi_{ar}} = \frac{\psi_{+-}\pi_{tr}}{\pi_{tr}+\pi_{ar}}\label{eq:PSBB}
\end{equation}
It is therefore equal to the Treatment Policy Estimand ($\psi_{+-}\pi_{tr}$) divided by the probability of being in $S_{+*}$, $\pi_{tr}+\pi_{ar}$. This probability can be estimated as $\hat{Pr}(Z=1|T=1)$, which means the estimand is very simple to estimate. A nice feature of this estimand is that it does not rely on the Monotonicity assumption for identification unlike the Treatment Policy Estimand in $S_{+-}$, because the proportion $\pi_{tr}$ does not need to be estimated on its own. 

\subsubsection{Simulation example}

To make things concrete, we report the results of each estimand strategy and subsequent sensitivity analysis when applied to 1000 simulated data sets of 500 individuals consistent with the DAG in Figure 8 (Top). A full summary of the data generating mechanism is given in Appendix B, but we give a few brief details below. The simulated data do not share {\it any} of the characteristics of the CANTOS trial.\\
\\
The prevalence of the binary outcome, $Y$, is 53\% across the simulated data. The mean proportion of biomarker responders in the $T=1$ group is 68\% and the mean proportion of biomarker responders in the $T=0$ group is 34\%. The normally distributed baseline covariate $X$ is a stronger predictor of biomarker response in the $T=1$ group than the $T=0$ group. The causal effect of biomarker response among those randomized to treatment is to reduce the outcome risk by 15\% ($\psi_{+*}=-0.15$). The causal effect of biomarker response among those randomized to control is to reduce the outcome risk by 5\% ($\psi_{++}=-0.05$). This means that the Homogeneity assumption is violated. All analyses follow the procedure outlined in Section 5.3 (with R,T, and S replaced with T, Z and X respectively) where average marginal effects are extracted from logistic regression model fits. Mean point estimates for the estimands are given in the upper half of Table \ref{tab:Sim1}. All estimates are obtained as Average Marginal Effects after fitting a logistic regression model with adjustment for measured confounder $X$. For completeness we also report the results of performing a naive `Responder analysis', which estimates the Average Marginal Effect between observed biomarker responders ($Z=1$) and non-responders ($Z=0$). 

\begin{table}[htbp]
\begin{center}
\begin{tabular}{l c c }
\hline
Estimand     &   Estimate    & Monte-Carlo SD  \\
\hline
&&\\
\multicolumn{3}{c}{{\bf Setting 1: Biomarker response}} \\
Policy                      &     -0.085 &  0.044 \\
Policy in S+-/Hypothetical  &    -0.250  &  0.130 \\
Responder Analysis          &   -0.059  &  0.045 \\
Policy in S+*               &   -0.130  &  0.065 \\
Hypothetical in S+*         &   -0.150  &  0.078 \\
&&\\

\multicolumn{3}{c}{{\bf Setting 2: Adherence}} \\

Policy        &   -0.37 &0.047 \\
Policy in S$_{++}$            &   -0.32 &0.038 \\
Policy in S$_{+*}$           &   -0.39 &0.051 \\
Per-Protocol  &   -0.26 &0.045 \\
$\alpha_{A}$      &   -0.40 &0.071 \\

\hline
\end{tabular}
\caption{\label{tab:Sim1} {\it Point estimates and Monte-Carlo standard deviations (SD) for the estimands considered in settings 1 and 2 across 1000 simulated data sets.}}
\end{center}
\end{table}

\subsection{Setting 2: Adjusting for general adherence}

\begin{figure}[htbp]
    \centering
\includegraphics[bb = 30 30 815 575, width=0.8\textwidth,clip]{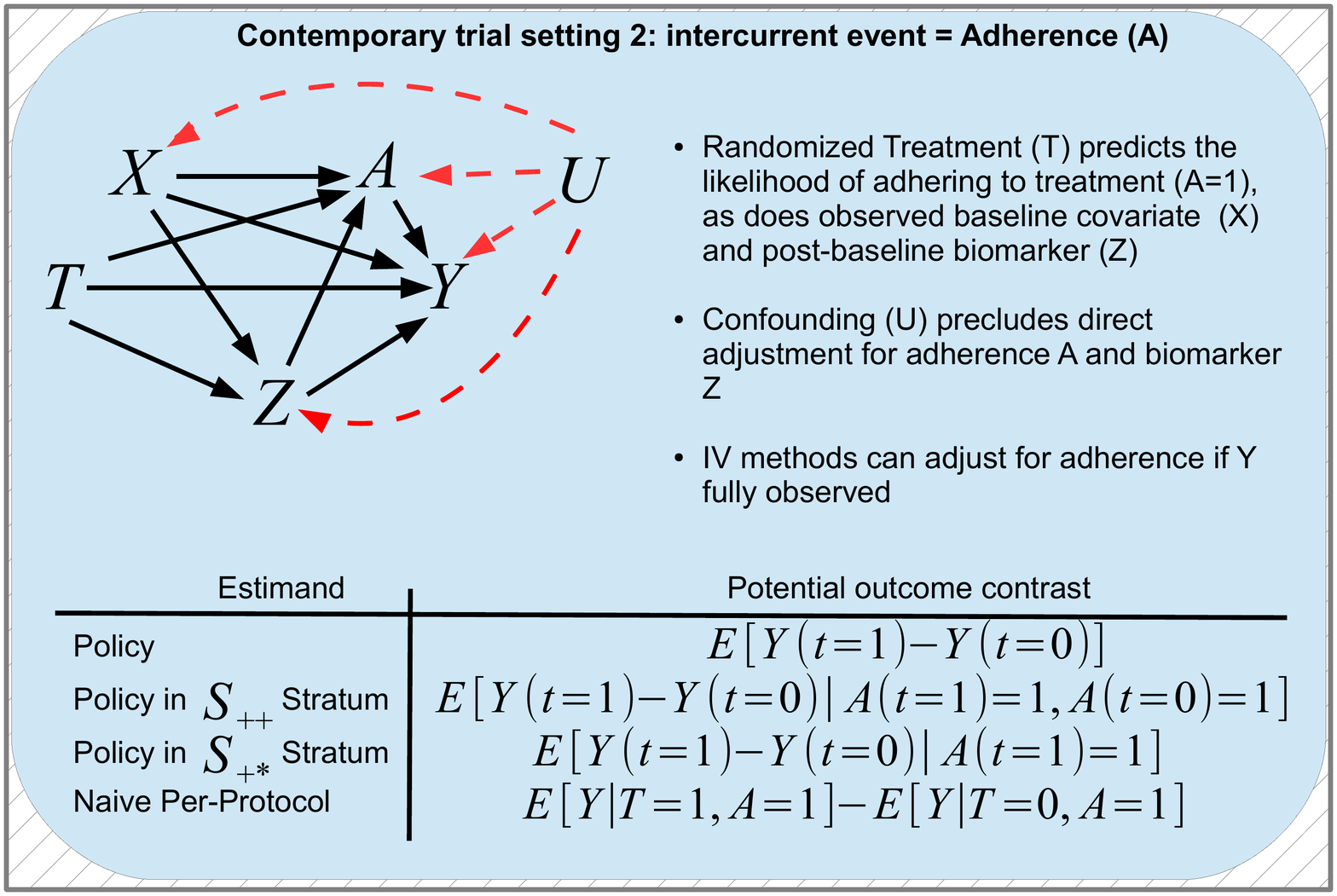}
\includegraphics[ width=0.55\textwidth,clip]{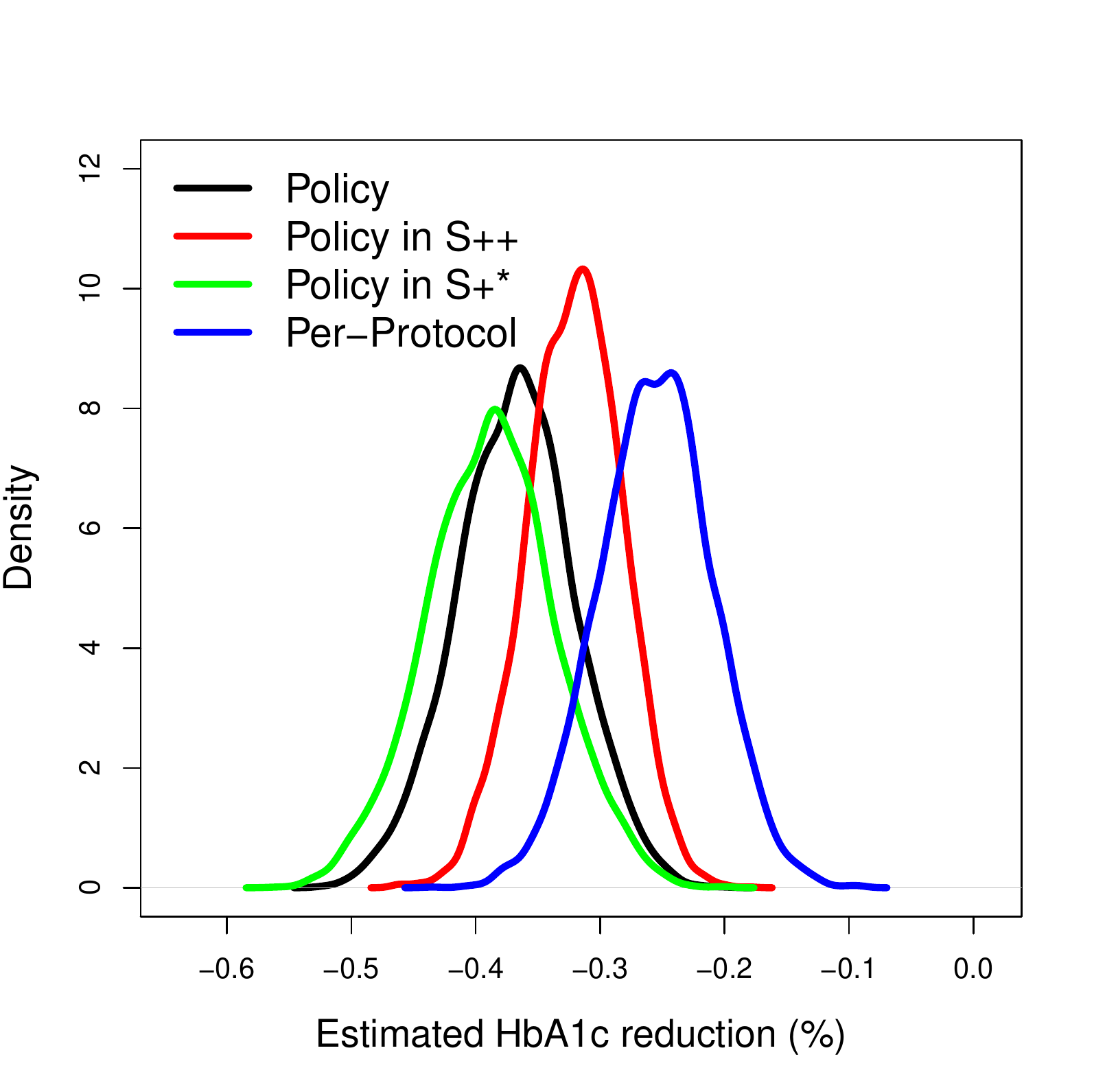}
\caption{{\it Top: An Instrumental Variable formulation contemporary industry trial setting 2: Intercurrent event = adherence. Bottom: Distribution of estimates for the simulation study of Section 6.2.2}}
\label{fig:estimandFigure2}
\end{figure}

We now consider a second example to clarify how IV methods can be used to adjust for general non-adherence.  In order to motivate ideas we use the data generating model consistent with the DAG in Figure 1 of Qu et. al. \cite{qu2020general}. In this setting randomized treatment $T$ exerts a direct effect on the outcome $Y$, but $T$, baseline covariate $X$ and post-baseline biomarker $Z$ all jointly predict the likelihood that an individual is adherent ($A=1$) or non-adherent ($A=0$) (Figure \ref{fig:estimandFigure2}, top). The variables $X$, $Z$, $A$ and $Y$ are all predicted by by a common unmeasured confounder, $U$.\\
\\
In this setting we have no interest in defining patient groups according to the biomarker $Z$, and assume that it is both unmeasured and ignored in subsequent analysis. Furthermore, even if we did measure $Z$ we may prefer not to adjust for it, because it could be predicted by unmeasured confounders and adjustment would then induce collider bias (see rule 3 Figure 1). Our interest instead lies in quantifying the Policy Estimand with within principled strata  defined by adherence status. In order to achieve this, we will assume that each individual is a member of one of four principal strata according to their adherence status $A(T)$ under assignment to treatment  $T=1$ and control $T=0$. Never Adherers ($na$) ($A(0)=A(1)=0$), or the $S_{--}$ stratum, wouldn't adhere to either treatment or control; Always Adherers ($aa$) ($A(0)=A(1)=1$), or the $S_{++}$ stratum, would adhere to both treatment or control; Treatment Adherers ($ta$) ($A(0)=0,A(1)=1$), or the $S_{+-}$ stratum, would adhere to Treatment but not control ; and Control Adherers ($ca$) ($A(0)=1,A(1)=0$), or the $S_{-+}$ stratum, would adhere to control but not treatment . Let $\pi_{na}$, $\pi_{aa}$, $\pi_{ta}$ and $\pi_{ca}$ represent the proportion of people in each principal strata. We assume the following model for the difference in potential outcomes for individual $i$ given $T=t,A=a$, and $T=t',A=a'$ :
\[\
E[Y_{i}(t,a)-Y_{i}(t',a')] = \psi(t-t') +  \alpha_{A}(a-a')
\]
This model implies that the Policy Estimand $E[Y_{i}(t=1)-Y_{i}(t=0)]$ in the $S_{--}$,  $S_{+-}$, $S_{++}$ and $S_{-+}$ strata are $\psi$, $\psi+\alpha_{A}$, $\psi$ and $\psi-\alpha_{A}$ respectively. By assuming the policy estimand in the $S_{--}$ and $S_{++}$ strata are non-zero, this structural model represents a full violation of the Exclusion Restriction, since:
\[\
E[Y_{i}(1,1)-Y_{i}(0,1)] = E[Y_{i}(1,0)-Y_{i}(0,0)] = \psi, 
\]
so that the equalities in Equation's (\ref{eq:ER1}) and (\ref{eq:ER2}) are violated. It also additionally assumes that the magnitude of the Exclusion Restriction violation, $\psi$, is the same. This means that the structural model is described by two parameters, which can both be estimated using the interaction technique previously described.

\subsubsection{Trial estimands and estimation}

Following Qu et al \cite{qu2020general} we can define 3 estimands of interest: the Treatment Policy Estimand in all subjects, the Policy Estimand within the Always Adherers ($S_{++}$), and the and the Policy Estimand within those who adhere to treatment ($A(1)$=1), which is the union of the $S_{++}$ and $S_{+-}$ strata. From the assumed potential outcome model we can see that the Policy Estimand in $S_{+*}$ is equal to

\begin{eqnarray}
&&\psi Pr(A(0)=1|A(1)=1) + (\psi+\alpha_{A}) Pr(A(0)=0|A(1)=1)  \nonumber \\
&=& \psi + \alpha_{A} Pr(A(0)=0|A(1)=1) \nonumber
\end{eqnarray}

To estimate the Policy Estimand in $S_{++}$ and $S_{+*}$ we therefore need a consistent estimate of $\psi$, $\alpha_{A}$ and $Pr(A(0)=0|A(1)=1)$.  To estimate $\psi$ and $\alpha_{A}$ we can again use the extended TSLS modelling procedure using $T$ and the interaction between $T$ and the baseline covariate $X$ as instruments to estimate the predicted adherence status of each individual, $\hat{A}$. The conditional probability term $Pr(A(0)=0|A(1)=1)$ can also be estimated from the data by 

\[\ 
\frac{\hat{Pr}(A=1|T=1)-\hat{Pr}(A=1|T=0)}{\hat{Pr}(A=1|T=1)}, \quad \text{ since } Pr(A(0)=0|A(1)=1)=\frac{\pi_{ta}-\pi_{ca}}{\pi_{ta}+\pi_{aa}} 
\]

Note that, as in the previous example, the Policy Estimand in the $S_{+*}$ stratum does not require the Monotonicity assumption for identification or estimation, but does require the probability of adhering to treatment to be higher on the $T=1$ arm than the $T=0$ arm, or equivalently that $\pi_{ta}-\pi_{ca}$ is greater than zero.

\subsubsection{Simulation example}

To make things concrete, we report the results of each estimand strategy and subsequent sensitivity analysis when applied to 1000 simulated data sets of 500 individuals consistent with the DAG in Figure \ref{fig:estimandFigure2} (Top) and the IMAGINE-3 study described by Qu et al. We note at the outset that our simulation is a simplification of that contained in Qu et al because we generate a single outcome for each patient rather than multiple repeated measures over time. A full summary of the data generating mechanism is given in Appendix C, but we give a few brief details below.\\
\\
The outcome, $Y$, is continuous with a mean value of 7.6 and a standard deviation of 0.7. It is intended to represent an HbA1c measurement on the \% scale. The mean proportion of adherers in treatment group is 70\% and the mean proportion of biomarker responders in the control group is 58\%. The normally distributed baseline covariate $X$ is a stronger predictor of adherence in the treatment group than the control group. The causal effect of treatment compared to control in all individuals is to reduce the HbA1c by 0.32\% ($\psi$=-0.32), which is itself the combination of a direct effect ($\psi_{t}$) of $-0.3$ and an indirect effect via $Z$ of $-0.02$\%. The additional effect of adherence, on either treatment, is to reduce HbA1c by a further 0.4\% ($\alpha_{A}=-0.4$). Although adherence modulates the outcome, we assume that all patient outcomes are observed. In this setting, implementing a two-parameter causal model, entails a first stage logistic regression of $A$ on $T$, $X$ and $T\times X$ to produce a predicted adherence variable $\hat{A}=Pr(A=1|T,X)$  (as in Section 5.3), followed by fitting a second stage linear regression model of the form 

\begin{equation}
Y_{i}|T_{i},X_{i} = \alpha_{0} + \psi T_{i}+ \alpha_{A}\hat{A}_{i} + \alpha_{X}X_{i} \nonumber 
\end{equation}

Mean point estimates for the estimands are given in the bottom half of Table \ref{tab:Sim1} and their distributions across the 1000 simulations are shown in Figure \ref{fig:estimandFigure2} (bottom). For completeness we also report the results of performing a naive `Per-Protocol analysis', which estimates the treatment effect only in the subgroup of individuals who adhered ($A=1$). All causal estimates closely agree with their theoretical estimands. The mean Policy estimate is -0.37. The mean Policy estimates in the $S_{++}$ and $S_{+*}$ strata are $-0.32$ and $-0.39$ respectively. The Per-Protocol estimate (i.e. the policy estimate among adherers) is $-0.26$\%. This is clearly biased, because conditioning directly on adherence status $A$ opens a biasing path from $T\rightarrow A \leftarrow U\rightarrow Y$, in this case leading to an underestimation of the effect.

\section{Discussion}
\label{s:disc}

In this paper we have attempted to explain the rationale for using Instrumental Variable methods in clinical trials. Starting from an academic trial perspective, we showed that when the intercurrent event is related to treatment adherence, then Policy, Hypothetical and standard Principal Stratum estimands can be estimated using a valid IV with the addition of either Monotonicity or Homogeneity. We described how a two parameter extension to the basic IV approach potentially allows the user to fit causal models that relax the Homogeneity and Exclusion restriction assumptions. Finally we showed how these methods can be applied to contemporary industry settings, where the intercurrent event is either a mechanistic consequence of treatment or some general measure of adherence. The two parameter approach is attractive, but requires the existence of baseline variables that strongly differentially predict the intercurrent event across treatment groups (which manifests as an interaction term in a regression model) and does not directly modulate the treatment effect. In practice, finding a baseline covariate that strongly and differentially predicts the intercurrent event without modulating the treatment effect may be challenging. Care must therefore be given when selecting a covariate for this role. Furthermore, if the interaction coefficient is small, large sample sizes will be needed to fit the two-parameter  models. This needs to be understood by trialists when planning the sample size of future trials if they wish to use these methods within the Estimand Framework.\\
\\
For a recent example where a such a plausible covariate exists, we refer to the AIRWAYS-2 trial \cite{benger2018}, which randomized paramedic teams to administer either a Tracheal intubation (TI) or a Supraglottic airway device (SGA) in out-of-hospital cardiac emergencies. In this trial it was impossible to receive the TI intervention, even if randomized to do so, if only one paramedic attended the scene in time (two or more were required). In follow up work, randomization and the interaction between randomization and the binary indicator variable $S$=I($\geq$ 2 paramedics attended) was deployed to fully adjust for non-adherence \cite{lazaroo2020}. In this case the interaction induced by $S$ was very strong but also unplanned. Future trialists  may consider building such a feature into the design in order to provide to facilitate the assessment of key assumptions within a causal analysis through the use of two-parameter models, as demonstrated here.\\
\\
In this paper we assumed that the intercurrent event of interest (e.g. adherence to treatment or biomarker response) was a binary variable. This makes it possible to apply the framework of Principal Stratification. In many cases this will be too simplistic a description, but it is in a sense actively encouraged by the Principal Stratification framework. However, this simplification is not inherent when defining Hypothetical estimands. For example, instead of dichotomising patients as biomarker responders/non-responders and non-responders, or treatment adherers/non-adherers, it would be possible to treat it as a continuous variable and use randomized treatment and baseline biomarker measurements to predict its value. A resulting Hypothetical estimand could then be constructed to reflect the difference in mean outcomes for all patients if their biomarker level had been lowered by a unit, with the choice of unit being user specified. In settings where the intercurrent event is non-adherence to the full dose of a treatment, Hypothetical estimands could also be constructed using information on the precise percentage of the treatment each patient took.\\
\\
Two estimand strategies mentioned in the E9 Addendum but not addressed in this paper are the so called `Composite Strategy' and the `While-on-Treatment Strategy'. Under the Composite Strategy, one can choose to integrate the intercurrent event as a component of the outcome variable itself in order to calculate the treatment effect. For a recent example of this, Permutt and Li \cite{permutt2017} investigated to address intercurrent events in a trial with a continuous outcome variable by assigning the missing outcomes a value lower than any observed value in the same arm. Outcomes are then ordered within each treatment group, equal proportions of data are trimmed away from each arm (the proportion being at least as large as the proportion of missing outcomes) and the treatment effect estimate is obtained using the trimmed data. Under the While-on-treatment Strategy the value of a patient response up to the time of the discontinuation of treatment (the intercurrent event) may be considered as a valid summary of their outcome. For example, Holzhauer et al. \cite{holzhauer2015} considered, amongst other estimands, treatment effects in a Diabetes setting up to the initiation of rescue medication to lower glucose levels. Dropout and discontinuation are of course ubiquitous in clinical trials, and one which the IV methods we have proposed do not account for. As future work, we will seek to  develop extended IV estimation strategies that can address this intercurrent event as well.

\appendix

\section{Data generating model for Section 5.4}

The randomization variable $R$, treatment variable $T$, baseline covariate $S$, outcome variable $Y$ and confounder variable $U$ for each subject was generated from the following model:

\begin{eqnarray}
R    &\sim& \text{Bern}(0.5) \nonumber \\
S    &\sim& \text{Bern}(0.5) \nonumber \\
U    &\sim& N(0,0.5)  \nonumber \\
\eta_{T} &=& -3 + 2R + 5RS + U, \quad P_{T} = \frac{\exp\left(\eta_{T}\right)}{1+\exp\left(\eta_{T}\right)} \nonumber \\
T    &\sim& \text{Bern}(P_{T}) \nonumber \\
Y    &=& 63 +\psi_{t}TR  +\psi_{at}T(1-R) + 3U  + 4\epsilon_{y},   \epsilon_{y} \sim N(0,3) \nonumber
\end{eqnarray}

Initially, we set $\psi_{t}$=20 and $\psi_{t}$=10 and then apply all approaches {\bf without} adjusting for the variable $U$. This produces the results shown in Figure \ref{fig:IVreg2} (left). We then set $\psi_{at}$=20 so that it equals $\psi_{t}$, and repeat the same analysis, but explicitly adjusting for $U$ in the outcome models. This produces the results shown in Figure \ref{fig:IVreg2} (right).

\section{Contemporary clinical trial setting 1: Biomarker response}

We used the following simulation model to generate data from an trial with heterogeneous treatment effects induced by biomarker response

\begin{eqnarray}
T        &\sim& \text{Bern}(0.5) \nonumber \\
U        &\sim& N(0,2) \nonumber \\
X        &\sim& -1 + U + N(n,0,2) \nonumber \\
\eta_{Z} &=& 0 + 3T + 2X - 4XT - 3U  \nonumber \\
P_{Z}    &=& \frac{\exp(\eta_{Z})}{1+\exp(\eta_{Z})}\nonumber \\
Z        &=& \text{Bern}(P_{Z})\nonumber \\
P_{Y}    &=& \alpha_{0} + \psi_{b}ZT +\psi_{ar}Z(1-T)  + \alpha_{X}X + \alpha_{U}U + \psi_{z}Z + 0.01N(0,1) \nonumber \\
Y        &=& \text{Bern}(P_{Y})\nonumber
\end{eqnarray}

In all analyses the parameter's $\alpha_{0}$, $\psi_{b}$, $\psi_{ar}$ and $\psi_{z}$ were set to 0.6, -0.15, -0.05 and 0.01 respectively.

\section{Contemporary clinical trial setting 2: General non-adherence}

We used the following simulation model to generate data from an adherence affected trial

\begin{eqnarray}
T        &\sim& \text{Bern}(0.5) \nonumber \\
U        &\sim& N(0,4) \nonumber \\
X        &\sim& 0.2U + N(n,0,2) \nonumber \\
Z        &=& 0.2X + 0.1U + 0.2T +N(0,1) \nonumber \\
\eta_{A} &=& 1 + 2T - 6XT + U + Z \nonumber \\
P_{A}    &=& \frac{\exp(\eta_{A})}{1+\exp(\eta_{A})}\nonumber \\
A        &=& \text{Bern}(P_{A})\nonumber \\
Y    &=& 8 + \psi_{t}T +\alpha_{A}A  - 0.1X + -0.1U + -0.1Z + N(0,0.2) \nonumber
\end{eqnarray}

The parameter $\psi_{t}$ was set to -0.3 and the parameter $\alpha_{A}$ was set to -0.4. The total treatment effect {\bf Not} through adherence, $\psi$, was equal to $\psi_{t} -0.1(E[Z|T=1]-E[Z|T=0])$ = -0.32  \\
\\

\verb R  code used to generate the illustrative trial data and perform the analyses discussed can be found in {\it Online Supplementary Methods}. 
\end{document}